             \documentstyle[12pt]{article}
\begin{document}\thispagestyle{empty}\begin{flushright}
OUT--4102--43\\
BI--TP/92--57                                       \end{flushright}
                                                    \vspace*{15mm}
                                                    \begin{center}{\large\bf
Two-loop two-point functions with masses:                     \\[4pt]
asymptotic expansions and Taylor series, in any dimension            }
                                                    \vglue 1cm{\bf
D.~J.~Broadhurst$^1$}
                                                    \vglue 3mm
Physics Department, Open University,\\
Milton Keynes, MK7 6AA, UK
                                                    \vglue 5mm{\bf
J.~Fleischer$^2$ and O.~V.~Tarasov$^3$}
                                                    \vglue 3mm
Fakult\"at f\"ur Physik, Universit\"at Bielefeld,\\
D-4800 Bielefeld 1, Germany                         \end{center}\vfill

{\bf Abstract} In all mass cases needed for quark and gluon self-energies,
the two-loop master diagram is expanded at large and small $q^2$, in $d$
dimensions, using identities derived from integration by parts.
Expansions are given, in terms of hypergeometric series, for all gluon
diagrams and for all but one of the quark diagrams; expansions of the
latter are obtained from differential equations. Pad\'{e} approximants to
truncations of the expansions are shown to be of great utility. As an
application, we obtain the two-loop photon self-energy, for all $d$, and
achieve highly accelerated convergence of its expansions in powers of
$q^2/m^2$ or $m^2/q^2$, for $d=4$.
                                            \vfill\begin{flushleft}
OUT--4102--43\\
BI--TP/92--57\\
April 1993                                  \end{flushleft}\vglue 5mm
                                            \footnoterule\noindent
$^1$) D.Broadhurst@open.ac.uk\\
$^2$) Fleischer@physik.uni-bielefeld.de\\
$^3$) Supported by Bundesministerium f\"ur Forschung and Technologie; on
leave of absence from the Joint Institute for Nuclear Research, 141980
Dubna, Russia
\newpage
\setcounter{page}{1}

\newcommand{\rd}{{\rm d}}
\newcommand{\ri}{{\rm i}}
\newcommand{\ep}{\varepsilon}
\newcommand{\Pm}{\phantom-}
\newcommand{\Pu}{\phantom1}
\newcommand{\Df}[2]{\mbox{$\frac{#1}{#2}$}}
\newcommand{\Pp}[3]{#1.#2\!\times\!10^{-#3}}
\newcommand{\Fh}[2]{\,{}_#1F_#2}
\newcommand{\Fs}[3]{\!\!\left[\begin{array}{c}#1\,;\\#2\,;\end{array}#3\right]}
\newcommand{\Fu}[2]{\Fs{#1}{#2}{1}}
\newcommand{\Ff}[2]{\Fs{#1}{#2}{4}}
\newcommand{\Fq}[2]{\Fs{#1}{#2}{\frac{1}{4}}}
\newcommand{\Fuq}[2]{\Fs{#1}{#2}{\frac{-q^2}{m^2}}}
\newcommand{\Fum}[2]{\Fs{#1}{#2}{\frac{m^2}{-q^2}}}
\newcommand{\Ffq}[2]{\Fs{#1}{#2}{\frac{-q^2}{4m^2}}}
\newcommand{\Ffm}[2]{\Fs{#1}{#2}{\frac{4m^2}{-q^2}}}
\newcommand{\Ffz}[2]{\Fs{#1}{#2}{z}}
\newcommand{\Ffu}[2]{\Fs{#1}{#2}{\frac{u}{4}}}
\newcommand{\Fft}[2]{\Fs{#1}{#2}{1-t}}
\newcommand{\Fzz}[2]{\Fs{#1}{#2}{1-z}}
\newcommand{\Li}[1]{\,{\rm Li}_{#1}}
\section{Introduction}

The agreement between LEP data~\cite{LEP} and standard-model predictions is
overwhelming~\cite{STATUS}. Deviations from the standard model would signal
{\em new\/} physics and are therefore intensively sought. For this reason,
high-precision standard-model radiative corrections must be computed,
entailing not only two-loop diagrams but also the effects of particle
masses. In many cases, radiative corrections to leading mass-dependent
effects are significant; in some cases~\cite{KK} they can reach 50\%. Thus,
in recent years, the attention devoted to two-loop diagrams with masses has
increased considerably~\cite{DJB}--\cite{DST}.

Whilst massless multi-loop diagrams can be dealt with by essentially
algebraic methods~\cite{CT}, the situation is more complex in the case of
massive diagrams. The first successful algebraic approach to massive
two-loop two-point functions achieved a systematic evaluation of diagrams
with external massive particles on the mass-shell~\cite{GBGS,SHELL2}, using
recurrence relations obtained from integration by parts in $d\equiv4-2\ep$
spacetime dimensions. These recurrence relations reduced a large variety of
integrals, depending on only one mass parameter, to combinations of
$\Gamma$\/-functions and a single so-called {\em master\/} integral that
contains all the problematic analytic properties of the Feynman diagrams.

As a next step, we investigate in this paper all the master integrals, with
one massive particle and an arbitrary external momentum, that occur in the
two-loop self-energy diagrams of QED and QCD, in an arbitrary dimension
$d$. Applications of the corresponding $d=4$ diagrams are made, for
example, in QCD sum rules~\cite{SVZ,GLUE}. There are several reasons for
evaluating them with arbitrary $d$.

First, it turns out~\cite{Q92} that $d$\/-dimensional calculations are
often much {\em easier\/} than their 4-dimensional counterparts. Recurrence
relations are best derived within the framework of dimensional
regularization, where contributions from surface terms in a partial
integration are disregarded. Moreover, these recurrence relations are
systematically implementable as computer-algebra algorithms. Hence the
algebraic methods of this paper are far less taxing than previous
4-dimensional analytical methods (as the author of~\cite{DJB} is only too
painfully aware).

Secondly, our methods yield not only two-loop results in
QED~\cite{KS,JS,BR}, QCD~\cite{PLB,SCG} and QCD$_2$~\cite{QCD2}, but are
also applicable to the dimensional regularization of {\em higher\/}-loop
calculations, where two-loop terms of order $\ep^{L-2}$ are needed to
obtain an $L$\/-loop result.

Thirdly, many of our results are analytically {\em simpler\/} than their
specializations to $d=4$. We find that the $d$\/-dimensional master
integrals are generally expressible in terms of simple hypergeometric
functions; only when these are expanded in $\ep$ do we encounter
complicated polylogarithms. {From} this point of view, the fearsome
analytic complexity of trilogarithms~\cite{LEW}, first encountered in the
two-loop QED calculation of~\cite{KS} and later surfacing in
QCD~\cite{PLB}, is merely a consequence of taking a singular limit of a few
otherwise very tractable hypergeometric functions of the type $\Fh21$ and
$\Fh32$.

Finally, the analytic continuation of hypergeometric functions giving a
Taylor series in $q^2/m^2$, to obtain those giving an asymptotic expansion
in $m^2/q^2$, is simple for arbitrary $d$, yet almost impossibly difficult
if one knows only the $d=4$ Taylor coefficients. Thus the {\em lesser\/}
labour of determining the $d$\/-dimensional small-$q^2$ expansion also
yields the {\em greater\/} benefit of a pair of complementary expansions,
to use for $d=4$.
\newpage

The objectives of this paper are as follows:
\begin{itemize}
\item[(a)] to obtain, whenever possible, hypergeometric representations
of the $d$\/-dimensional two-loop master integrals of QED and QCD;
\item[(b)] when this appears impossible, to obtain the differential
equation satisfied by the $d$\/-dimensional master integral;
\item[(c)] to generate, thereby, any desired number of terms in
both the small-$q^2$ Taylor series and also the
large-$q^2$ asymptotic expansion, for any $d$;
\item[(d)] to demonstrate the utility of applying methods of accelerated
convergence to each of the two series thereby obtained;
\item[(e)] to exemplify these algebraic and numerical methods
in the concrete case of the self-energy of the photon.
\end{itemize}

These aims are complementary to those of~\cite{DST}, which extends the
calculation of leading terms in the asymptotic expansion, with $d=4$, to
mass cases more complicated than those considered in~\cite{DJB}. Here we
restrict attention to the mass cases of~\cite{DJB}, whilst both
significantly generalizing and also greatly simplifying the analysis, by
working with arbitrary $d$. Moreover, the knowledge we derive from the
complete Taylor series and from the complete asymptotic expansion enables
us to investigate methods of accelerated convergence~\cite{EPS}, which we
find to be of great utility. We therefore proceed as follows.

In Section~2 we consider all the mass cases of~\cite{DJB}, for arbitrary
$d$. These are illustrated in Fig.~1. For all cases except that of
Fig.~1(c), we achieve objectives~(a) and~(c), above. The irreducibility to
simple hypergeometric series of integral $I_3$ of Fig.~1(c) is to be
expected; for $d=4$ it is not reducible to polylogarithms, being rather a
function whose discontinuity has a derivative involving an elliptic
integral~\cite{DJB}. For $I_3$, we achieve objectives~(b) and~(c), using
differential equations of the type developed in~\cite{DJB,KOT}.
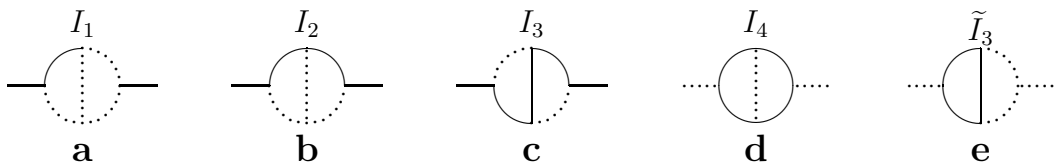
\begin{center}{\bf Fig.~1:} The mass cases of~\cite{DJB}:
dotted lines are massless; solid lines have mass $m$.\end{center}
\setlength{\unitlength}{0.01cm}
\newbox\shell
\def\pxl{\circle*{3}}
\def\Ss{\scriptstyle}
\def\align#1#2{
 \dimen0=\ht\shell
 \multiply\dimen0 by #1
 \divide\dimen0 by #2
 \raise-\dimen0\box\shell}
\def\arca{
 \put(11.1,48.7)\pxl
 \put(21.7,45.0)\pxl
 \put(31.2,39.1)\pxl
 \put(39.1,31.2)\pxl
 \put(45.0,21.7)\pxl
 \put(48.7,11.1)\pxl
 \put(50.0, 0.0)\pxl}
\def\arcb{
 \put(48.7,-11.1)\pxl
 \put(45.0,-21.7)\pxl
 \put(39.1,-31.2)\pxl
 \put(31.2,-39.1)\pxl
 \put(21.7,-45.0)\pxl
 \put(11.1,-48.7)\pxl
 \put( 0.0,-50.0)\pxl}
\def\arcc{
 \put(-11.1,-48.7)\pxl
 \put(-21.7,-45.0)\pxl
 \put(-31.2,-39.1)\pxl
 \put(-39.1,-31.2)\pxl
 \put(-45.0,-21.7)\pxl
 \put(-48.7,-11.1)\pxl
 \put(-50.0, 0.0)\pxl}
\def\arcd{
 \put(-48.7, 11.1)\pxl
 \put(-45.0, 21.7)\pxl
 \put(-39.1, 31.2)\pxl
 \put(-31.2, 39.1)\pxl
 \put(-21.7, 45.0)\pxl
 \put(-11.1, 48.7)\pxl
 \put( 0.0, 50.0)\pxl}
\hfill
\setbox\shell=\hbox{
 \begin{picture}(200,200)(-100,-100)
  \put(0,0){\oval(100,100)[tl]}
  \put(-50,0){\line(-1,0){50}}
  \put( 50,0){\line( 1,0){50}}
  \multiput(0,50)(0,-10){10}\pxl
 \arca\arcb\arcc
 \put(-100,-100){\makebox(200,200)[t]{$I_1$}}
 \put(-100,-100){\makebox(200,200)[b]{\large\bf a}}
 \end{picture}}
\align{7}{16}\hfill
\setbox\shell=\hbox{
 \begin{picture}(200,200)(-100,-100)
 \put(0,0){\oval(100,100)[t]}
 \put(-50,0){\line(-1,0){50}}
 \put( 50,0){\line( 1,0){50}}
 \multiput(0,47)(0,-10){10}\pxl
 \arcb\arcc
 \put(-100,-100){\makebox(200,200)[t]{$I_2$}}
 \put(-100,-100){\makebox(200,200)[b]{\large\bf b}}
 \end{picture}}
\align{7}{16}\hfill
\setbox\shell=\hbox{
 \begin{picture}(200,200)(-100,-100)
 \put(0,0){\oval(100,100)[tr]}
 \put(0,0){\oval(100,100)[bl]}
 \put(-50,0){\line(-1,0){50}}
 \put( 50,0){\line( 1,0){50}}
 \put( 0,50){\line(0,-1){100}}
 \arcb\arcd
 \put(-100,-100){\makebox(200,200)[t]{$I_3$}}
 \put(-100,-100){\makebox(200,200)[b]{\large\bf c}}
 \end{picture}}
\align{7}{16}\hfill
\setbox\shell=\hbox{
 \begin{picture}(200,200)(-100,-100)
 \put(0,0){\circle{100}}
 \multiput(-55,0)(-10,0){5}\pxl
 \multiput( 55,0)( 10,0){5}\pxl
 \multiput(0,47)(0,-10){10}\pxl
 \put(-100,-100){\makebox(200,200)[t]{$I_4$}}
 \put(-100,-100){\makebox(200,200)[b]{\large\bf d}}
 \end{picture}}
\align{7}{16}\hfill
\setbox\shell=\hbox{
 \begin{picture}(200,200)(-100,-100)
 \put(0,0){\oval(100,100)[l]}
 \multiput(-55,0)(-10,0){5}\pxl
 \multiput( 55,0)( 10,0){5}\pxl
 \put( 0,50){\line(0,-1){100}}
 \arca\arcb
 \put(-100,-100){\makebox(200,200)[t]{$\widetilde{I}_3$}}
 \put(-100,-100){\makebox(200,200)[b]{\large\bf e}}
 \end{picture}}
\align{7}{16}\hfill\mbox{\hspace{0.5cm}}\vskip0.2cm

In Section~3 we study the $d\to4$ and $d\to3$ limits of the most difficult
integral, $I_3$, whose Taylor series diverges for $|q^2|>m^2$, whilst its
asymptotic expansion diverges for $|q^2|<9m^2$. A method for separating the
sources of discontinuity is given in Section~3.1, allowing series
expansions throughout the entire complex plane. This is shown to entail
working to high numerical precision at intermediate stages of the
computations.

In Section~3.2 we find a more efficient solution to the problem of
intermediate space-like values of $q^2$, namely the application of Pad\'{e}
approximants to the Taylor series, and their extension, via the so-called
$\ep$\/-algorithm~\cite{EPS}, to achieve accelerated convergence of the
asymptotic expansion. The improvement of the convergence of the expansions,
within their respective domains of validity, is impressive. Even more
significant is the high numerical accuracy achieved in the space-like region
$q^2\in[m^2,9m^2]$, where {\em neither\/} of the original series was valid.
This extension of the domain of convergence is known to occur when one
applies Pad\'{e} approximants to the Taylor series of a function with a
positive spectral density, namely to a function of the so-called Stieltjes
type~\cite{EPS}. We have no right to expect it to occur for the asymptotic
expansion, since that involves both powers and logarithms of $m^2/q^2$,
when $d=4$. We regard the numerical results of Section~3.2 as highly
significant; such dual application of methods of accelerated convergence to
small-$q^2$~\cite{INLO} and large-$q^2$~\cite{DST} series may well be the
next step to take in extending the scope and accuracy of practical
standard-model calculations.

In Section~4 we apply the methods of previous sections to the photon
self-energy. First we show how analytically simple are the new
$d$\/-dimensional results for the self-energy and its spectral density.
Spin is truly an inessential complication in $d$ dimensions; the methods
that yield the corresponding master integral also yield the full
self-energy in terms of precisely three simple hypergeometric functions.
Secondly, we show how impressively accurate are the results of accelerating
the convergence of the Taylor series and the asymptotic expansion.

In Section~5 we summarize our findings and draw conclusions from the
efficiency of our algebraic methods and the accuracy of our numerical
approximations.

\section{Master integrals}

We consider the generalization to $d=4-2\ep$ space-time dimensions of
integrals analyzed in~\cite{DJB} in the case $d=4$, namely the (euclidean)
integrals
\begin{eqnarray}
I_0&=&\int\!\!\int\frac{\rd^dk_1\rd^dk_2}{\pi^d\Gamma^2(1+\ep)}
P(k_1,0)P(k_2,0)P(k_1-q,0)P(k_2-q,0)P(k_1-k_2,0),
\label{i0}\\
I_1&=&\int\!\!\int\frac{\rd^dk_1\rd^dk_2}{\pi^d\Gamma^2(1+\ep)}
P(k_1,m)P(k_2,0)P(k_1-q,0)P(k_2-q,0)P(k_1-k_2,0),
\label{i1}\\
I_2&=&\int\!\!\int\frac{\rd^dk_1\rd^dk_2}{\pi^d\Gamma^2(1+\ep)}
P(k_1,m)P(k_2,m)P(k_1-q,0)P(k_2-q,0)P(k_1-k_2,0),
\label{i2}\\
I_3&=&\int\!\!\int\frac{\rd^dk_1\rd^dk_2}{\pi^d\Gamma^2(1+\ep)}
P(k_1,0)P(k_2,m)P(k_1-q,m)P(k_2-q,0)P(k_1-k_2,m),
\label{i3}\\
I_4&=&\int\!\!\int\frac{\rd^dk_1\rd^dk_2}{\pi^d\Gamma^2(1+\ep)}
P(k_1,m)P(k_2,m)P(k_1-q,m)P(k_2-q,m)P(k_1-k_2,0),
\label{i4}\\
\widetilde{I}_3&=&\int\!\!\int\frac{\rd^dk_1\rd^dk_2}{\pi^d\Gamma^2(1+\ep)}
P(k_1,m)P(k_2,0)P(k_1-q,m)P(k_2-q,0)P(k_1-k_2,m),
\label{i3t}
\end{eqnarray}
where $P(k,m)\equiv1/(k^2+m^2)$ and the subscript of $I_N$ denotes the
number of massive lines. The denominator structure of
integrals~(\ref{i1}--\ref{i3t}) is illustrated in Fig.~1. (Note that the
$d\to4$ limits of these integrals, given in~\cite{DJB}, differ by a factor
of $q^2$.) Integrals $\{I_0,\widetilde{I}_3,I_4\}$ are needed for the
two-loop gluon self-energy; integrals $\{I_2,I_3\}$ for the quark
self-energy. As in~\cite{DJB}, we include the integral $I_1$ for
illustrative purposes.

\subsection{Integral $I_1$: an example of Mellin--Barnes techniques}

With only one massive line, the integral~(\ref{i1}) of Fig.~1(a) is most
easily evaluated by using the Mellin--Barnes representation~\cite{JMP}
\begin{equation}
P(k_1,m)\equiv\frac{1}{k_1^2+m^2}=\frac{1}{2\pi\ri\,k_1^2}
\int\limits_{-\frac12-\ri\infty}^{-\frac12+\ri\infty}\rd s\,
\Gamma(-s)\Gamma(1+s)\left[\frac{m^2}{k_1^2}\right]^{\displaystyle s},
\label{mb}
\end{equation}
where the contour separates the poles of the integrand on the right, with
${\rm Re}\,s\geq0$, from those on the left, with ${\rm Re}\,s\leq-1$.
Inserting~(\ref{mb}) in~(\ref{i1}), one obtains an integral over $s$ whose
integrand is given in terms of $\Gamma$-functions by the triangle rule
of~\cite{CT}. Closing the contour to the left, we find the small-$q^2$
expansion
\begin{eqnarray}
q^2m^{4\ep}\ep^3(1-2\ep)I_1&=&
\Df12\delta\Fh21\Fuq{\ep,2\ep}{1-\ep}
+(\beta-\Df12\delta)\Fh21\Fuq{1,\ep}{1-\ep}
\nonumber\\&&{}
+\Df12\gamma\Fh21\Fuq{1,-\ep}{1-3\ep}-\beta-\Df12\gamma,
\label{i1q}
\end{eqnarray}
where
\begin{eqnarray}
\beta&\equiv&\frac{\Gamma^2(1-\ep)}{\Gamma(1-2\ep)}
\left(\frac{m^2}{q^2}\right)^\ep=1-\ep\ln(q^2/m^2)+O(\ep^2),
\label{bet}\\
\gamma&\equiv&\frac{\Gamma(1+2\ep)}{\Gamma^2(1+\ep)}
\frac{\Gamma^3(1-\ep)}{\Gamma(1-3\ep)}
\left(\frac{m^2}{q^2}\right)^{2\ep}=\beta^2-6\zeta(3)\ep^3+O(\ep^4),
\label{gam}\\
\delta&\equiv&\frac{\Gamma(1-\ep)\Gamma(1+2\ep)}{\Gamma(1+\ep)}
=1+2\zeta(2)\ep^2+O(\ep^3)
\label{del}
\end{eqnarray}
are structures that will recur in our analysis.
Closing the contour to the right, we find the large-$q^2$ expansion
\begin{eqnarray}
&&q^2m^{4\ep}\ep^3(1-2\ep)I_1=
-\Df12\delta
+\beta^2\left(1+\frac{m^2}{q^2}\right)^{-2\ep}
+(\Df12\delta-\beta)\Fh21\Fum{1,\ep}{1-\ep}
\nonumber\\&&{}
+\beta\Fh21\Fum{\ep,2\ep}{1-\ep}
-\Df12\gamma\Fh21\Fum{2\ep,3\ep}{1+\ep}
-\Df12\gamma\Fh21\Fum{1,3\ep}{1+\ep},
\label{i1m}
\end{eqnarray}
which may also be obtained by inverting the hypergeometric functions
of~(\ref{i1q}).

As $d\to4$, we find that the same function appears in the small-$q^2$
expansion~(\ref{i1q}) and the large-$q^2$ expansion~(\ref{i1m}):
\begin{equation}
\left.q^2I_1\right|_{d=4}=-F_1(q^2/m^2)=6\zeta(3)+F_1(m^2/q^2),
\label{i1qm}
\end{equation}
where
\begin{equation}
F_1(x)=\sum_{n=1}^\infty\left(\frac{\ln^2x+2\zeta(2)}{2n}
-\frac{2\ln x}{n^2}+\frac{3}{n^3}
+2\sum_{r=1}^{n-1}\frac{n-r}{r^2n^2}\right)(-x)^n,
\label{f1x}
\end{equation}
for $|x|<1$, in agreement with Eq.~(14) of~\cite{DJB}.

\subsection{Integral $I_4$: for abelian boson self-energies}

The small-$q^2$ expansion of integral~(\ref{i4}) of Fig.~1(d) is obtained
by extending the method of integration by parts, used in~\cite{CT} for the
massless integral~(\ref{i0}). The triangle rule eliminates
either the massless line or two of the massive lines. In the former case,
one has the product of two one-loop integrals; in the latter case, the
external momentum may be routed through the massless line, making the
small-$q^2$ expansion straightforward.
In terms of hypergeometric functions that
are regular as $q^2\to0$, the $d$-dimensional result is
\begin{eqnarray}
&&m^{2+4\ep}\ep(1-2\ep)I_4=
\left(1+\frac{q^2}{4m^2}\right)G_4^2-\frac{H_4}{(1+2\ep)(1-\ep)},
\label{i4q}\\
&&G_4\equiv\Fh21\Ffq{1,1+\ep}{\frac32},\quad
H_4\equiv\Fh32\Ffq{1,1+\ep,1+2\ep}{\frac32+\ep,2-\ep},
\label{gh4}
\end{eqnarray}
which gives the small-$q^2$ expansion in Eq.~(44) of~\cite{DJB}, as $d\to4$.

This result may be transformed into one involving hypergeometric series
that are regular as $q^2\to\infty$, by inverting the series~(\ref{gh4}), to
obtain:
\begin{eqnarray}
&&q^2m^{4\ep}\ep^3(1-2\ep)I_4=
\left(1+\frac{4m^2}{q^2}\right)^{-2\ep}\left(\beta
-\Fh21\Ffm{\frac12-\ep,-\ep}{1-\ep}\right)^2
\phantom{push left to clear}
\nonumber\\&&{}
-\gamma\Fh21\Ffm{\frac12+\ep,3\ep}{1+\ep}
+2\beta\Fh21\Ffm{\frac12,2\ep}{1-\ep}
-\Fh32\Ffm{1,\frac12-\ep,\ep}{1-\ep,1-2\ep},
\label{i4m}
\end{eqnarray}
from which the large-$q^2$ expansion of Eq.~(44) of~\cite{DJB} is
obtained in the limit $d\to4$.

\subsection{Integral $I_2$: for non-abelian fermion self-energies}

The key to obtaining results~(\ref{i4q},\ref{i4m}) for $I_4$ was the
generalization of the triangle rule of~\cite{CT}, which applies when a
massless particle is exchanged between particles whose masses do not
change~\cite{KOT}. This method is therefore applicable to
integral~(\ref{i2}) of Fig.~1(b), for which one obtains the small-$q^2$
expansion
\begin{equation}
m^{2+4\ep}\ep^2(1-\ep)(1-2\ep)I_2=
\delta H_2-G_2+\frac{\ep}{1-\ep}\frac{q^2}{m^2}G_2^2\,,
\label{i2q}
\end{equation}
with $\delta$ given by~(\ref{del}) and
\begin{equation}
G_2\equiv\Fh21\Fuq{1,1+\ep}{2-\ep},\quad
H_2\equiv\Fh21\Fuq{1+\ep,1+2\ep}{2-\ep}.
\label{gh2}
\end{equation}
Inverting the hypergeometric series, we obtain the large-$q^2$ expansion
\begin{equation}
q^2m^{4\ep}\ep^3(1-2\ep)I_2=
\beta\Fh21\Fum{1+\ep,2\ep}{1-\ep}
-\gamma\Fh21\Fum{1+2\ep,3\ep}{1+\ep}-F_2+F_2^2,
\label{i2m}
\end{equation}
with
\begin{equation}
F_2\equiv\Fh21\Fum{1,\ep}{1-\ep}-\beta\left(1+\frac{m^2}{q^2}\right)^{-2\ep}
=\frac{\ep}{1-\ep}\frac{q^2}{m^2}G_2\,.
\label{f2}
\end{equation}
The expansions~(\ref{i2q},\ref{i2m}) give the coefficients of Eq.~(22)
of~\cite{DJB}, as $d\to4$.

\subsection{Integral $\widetilde{I}_3$: for non-abelian boson
self-energies}

More difficult than either of the integrals~(\ref{i2},\ref{i4}) is
integral~(\ref{i3t}) of Fig.~1(e). In this case, the triangle rule fails to
simplify the integral, since the exchanged particle is massive.
Nevertheless, the small-$q^2$ expansion of a combination of the integral
and its derivative w.r.t.\ $q^2$ can be reduced to $\Fh32$ and $\Fh21$
functions, using integration by parts. {From} this result, an expansion in
terms of $\Fh43$ and $\Fh32$ functions is obtained by integration, which
yields
\begin{eqnarray}
&&m^{2+4\ep}(1+\ep)(1-2\ep)\widetilde{I}_3=\frac{1}{(1-\ep)(1+2\ep)}
\Fh43\Ffq{1,1+\ep,1+\ep,1+2\ep}{\frac32+\ep,2+\ep,2-\ep}
\nonumber\\
&&\quad{}+\frac{1+\ep}{2\ep}\beta\Fh32\Ffq{1,1,1+\ep}{\frac32,2}
 -\frac{1}{2\ep}\Fh32\Ffq{1,1+\ep,1+\ep}{\frac32,2+\ep}.
\label{i3tq}
\end{eqnarray}
The presence of $(m^2/q^2)^\ep$, via $\beta$ in the second term
of~(\ref{i3tq}), reveals the branchpoint at $q^2=0$, which prohibits the
development of a simple Taylor series. Instead, one obtains
\begin{equation}
\widetilde{I}_3=\frac{1}{2m^2}\sum_{n=0}^\infty
\frac{n!\,\Gamma(\Df32)}{\Gamma(n+\Df32)}
\left(\frac{-q^2}{4m^2}\right)^n\left(\frac{3}{(n+1)^2}
-\frac{\ln(q^2/m^2)}{n+1}\right)+O(\ep),
\label{i3t4}
\end{equation}
in agreement with Eq.~(45) of~\cite{DJB}.

The inversion of~(\ref{i3tq}), to produce a large-$q^2$ expansion, is
likewise difficult, since each hypergeometric function has a repeated
argument. To outflank this difficulty, one returns to the differential
equation relating the integral and its derivative w.r.t.\ $m^2$ to simpler
hypergeometric functions, whose inversion is not frustrated by repeated
arguments. Inverting the latter and integrating the resultant series, one
obtains
\begin{eqnarray}
&&q^2m^{4\ep}\ep^3(1-2\ep)\widetilde{I}_3=
\Fh43\Ffm{1,\frac12-\ep,\ep,-\ep}{1-\ep,1-\ep,1-2\ep}
-\gamma\Fh32\Ffm{\frac12+\ep,\ep,3\ep}{1+\ep,1+\ep}
\nonumber\\&&{}
+\beta^2\Fh21\Ffm{\frac12+\ep,\ep}{1+\ep}
-\Fh32\Ffm{1,\frac12,-\ep,}{1-\ep,1-\ep}
+\left(\frac{2\ep}{1-\ep}
\Fh32\Ffm{1,1,\frac32}{2,2-\ep}
\right.\nonumber\\
&&\left.{}+2\ep(1+2\ep)
\Fh32\Ffm{1,1,\frac32+\ep}{2,2}
-\frac{8\ep^2}{1-\ep}
\Fh43\Ffm{1,1,\frac32,1+2\ep}{2,2,2-\ep}
\right)\frac{\beta m^2}{q^2}.
\label{i3tm}
\end{eqnarray}
The $d\to4$ limit agrees with Eq.~(45) of~\cite{DJB}; the $m\to0$ limit
agrees with the massless result of~\cite{CT} for integral~(\ref{i0}):
\begin{equation}
I_0=\frac{q^{2(d-5)}}{1-2\ep}\frac{\Gamma(1+2\ep)}{\Gamma^2(1+\ep)}
\frac{\Gamma^3(-\ep)}{\Gamma(1-3\ep)}
\left(1-\frac{\Gamma(1+\ep)\Gamma(1-3\ep)}{\Gamma(1-2\ep)}
\cos\pi\ep\right)=\frac{6\zeta(3)}{q^2}+O(\ep),
\label{i0m}
\end{equation}
which is likewise obtained from~(\ref{i1m},\ref{i4m},\ref{i2m}).

\subsection{Integral $I_3$: for abelian fermion self-energies}

The remaining integral~(\ref{i3}) of Fig.~1(c) is by far the most
difficult. Symptomatic of its intractability are the features~\cite{DJB}
found for $d=4$, where the small-$q^2$ expansion involves the maximum value
of Clausen's integral, which is not reducible to $\zeta$-functions, and
the derivative of the discontinuity in the time-like region
$-q^2>9m^2$ is an elliptic integral. Accordingly, we generalize the method
of~\cite{DJB} to arbitrary $d$, obtaining differential equations whose
right-hand sides do not involve intermediate states with
three massive particles.

First we treat the simplest two-point function with three massive propagators:
\begin{equation}
J_3\equiv\int\!\!\int\frac{\rd^dk_1\rd^dk_2}{\pi^d\Gamma^2(1+\ep)}
P(k_2,m)P(k_1-q,m)P(k_1-k_2,m),
\label{j3}
\end{equation}
for which a Mellin--Barnes transform of one propagator yields
\begin{equation}
-m^{4\ep-2}\ep^2(1-\ep)(1-2\ep)J_3=
\sum_{n=0}^\infty\frac{A(n,q^2)+(1-2\ep)B(n,q^2)}{4^n}\,,
\label{j3mb}
\end{equation}
\vspace{-0.3cm}
\begin{eqnarray}
A(n,q^2)&=&\Fh32\Ffq{n+1,n+\ep,n-1+2\ep}{2-\ep,n+\frac12+\ep}
\frac{(-1+2\ep)_n}{(\frac12+\ep)_n},
\label{anh}\\
B(n,q^2)&=&\Fh32\Ffq{n+1,n+\ep,n+2-\ep}{2-\ep,n+\frac32}
\frac{(\ep)_n}{(\frac32)_n},
\label{bnh}
\end{eqnarray}
with $(a)_n\equiv\Gamma(a+n)/\Gamma(a)$.
At $q^2=-m^2$, we obtain from~(\ref{j3mb}--\ref{bnh}) the on-shell result
\begin{eqnarray}
\!\!\!\left.-m^{4\ep-2}\ep^2(1-\ep)(1-2\ep)J_3\right|_{q^2=-m^2}
\!\!\!\!\!&=&\!\!\!\Fh32\Fu{1,-1+2\ep,\frac12}{\frac12+\ep,2-\ep}
+(1-2\ep)\Fh32\Fu{1,\ep,\frac32-\ep}{\frac32,3-2\ep}
\nonumber\\
&=&\!\!\Df32-\Df14\ep-\Df{19}{8}\ep^2+O(\ep^3)\,,
\end{eqnarray}
whose $\ep$-expansion may extended, through $O(\ep^6)$,
in terms of $\{{\rm Li}_n(1),{\rm Li}_n(\frac12)|n\leq5\}$~\cite{3LP}.

After some transformations,~(\ref{j3mb}) can be rewritten as
\begin{equation}
-{m}^{4\ep-2}{\ep}^{2}(1-\ep)(1-2\ep)J_3
=\sum_{n=0}^{\infty}C(n)\frac{(-q^2/m^2)^n}{(2-\ep)_n n!}\,,
\label{j3q}
\end{equation}
\vspace{-0.3cm}
\begin{equation}
C(n)=\frac{\rd^n}{\rd u^n}\Bigg\{
u^{n-1+\ep}\frac{\rd^n}{\rd u^n}\Fh21\Ffu{1,-1+2\ep}{\frac12+\ep}
+(1-2\ep)u^{n+1-\ep}\frac{\rd^n}{\rd u^n}\Fh21\Ffu{1,\ep}{\frac32}
\Bigg\}\Bigg|_{u=1}.
\label{cn}
\end{equation}
For the leading term in~(\ref{j3q}), we obtain from~(\ref{cn}) the
vacuum-diagram result~\cite{INLO,FJ}:
\begin{equation}
C(0)=\frac{\pi}{3^{\ep-\frac12}}\,\frac{\Gamma(2\ep)}{\Gamma^2(\ep)}
+\Df32(1-2\ep)\Fh21\Fq{1,\ep}{\frac32}
=\Df32(1-9\ep^2S_2)+O(\ep^3),
\label{j3z}
\end{equation}
where $S_2$~\cite{DJB} is a multiple of ${\rm Cl}_2(\pi/3)$,
the maximum value of Clausen's integral~\cite{LEW}:
\begin{eqnarray}
S_2&\equiv&\sum_{n=1}^\infty\frac{2n-1}{(3n-1)^2(3n-2)^2}
=-\frac{4}{9\sqrt{3}}\int_0^{\pi/3}\rd\theta
\ln\left(2\sin\frac{\theta}{2}\right)
\label{s2}\\
&=&0.260\,434\,137\,632\,162\,098\,955\,729\,143\,208\,\ldots
\label{s2v}
\end{eqnarray}
and is needed to high precision, for the numerical analysis of Section~3.1.
It is tedious to obtain higher coefficients from~(\ref{cn}).
It is easier to use the second-order differential equation that follows
from~(\ref{anh},\ref{bnh}), namely
\begin{eqnarray}
&&2(q^2+m^2)(q^2+9m^2)m^2J_3^{\prime\prime}
-\left((d-4)q^4+10(3d-10)q^2m^2+9(5d-16)m^4\right)J_3^\prime
\nonumber\\&&{}
+3(3d-8)(d-3)(q^2+3m^2)J_3
=\frac{48q^2m^{2d-6}}{(d-4)^2}\,,
\label{jde}
\end{eqnarray}
where primes denote differentiation w.r.t.\ $m^2$. This differential
equation may also be obtained directly, using integration by parts. It
yields the recurrence relation
\begin{eqnarray}
C(n+1)&=&\Df13(d-2)(d-3)\delta_{n,0}
-\Df{1}{18}\left\{(d+4)(d-3)+10n(d-2n-4)\right\}C(n)
\nonumber\\&&{}
-\Df{1}{36}n(d+2n-2)(d-2n-2)(d-n-2)C(n-1)\,,
\label{j3rec}
\end{eqnarray}
which enables efficient computation of any desired number of small-$q^2$
coefficients.

The expansion at large $q^2$ is of the form
\begin{equation}
-q^2m^{4\ep}\ep^2(1-2\ep)J_3=
\sum_{n=0}^\infty\left(\gamma q^4C_1(n)+\beta q^2m^2C_2(n)+m^4C_3(n)\right)
\frac{(-m^2/q^2)^n}{n!},
\label{j3m}
\end{equation}
where $C_1(0)=\ep/\left[2(1-3\ep)(2-3\ep)\right]$, $C_2(0)=3/(1-\ep)$, and
the remaining coefficients are rational functions of $\ep$, determined
by~(\ref{jde}). These coefficients may be obtained in closed form,
by transforming the hypergeometric series~(\ref{anh},\ref{bnh}) into
large-$q^2$ expansions. This inversion generates the
structures~(\ref{bet},\ref{gam}). Collecting powers of $-m^2/q^2$,
we obtain the terminating series
\begin{eqnarray}
C_1(n)&=&
\frac{(-1+2\ep)_n(3\ep)_{n-2}}{2(1+\ep)_{n-1}}
\Fh32\Ff{-\frac12+\ep,1-\ep-n,-n}{-1+2\ep,\ep},
\label{c1}\\
C_2(n)&=&
\frac{(-1+2\ep)_n(\ep)_n}{(1-\ep)(2-\ep)_n}
\Fh32\Ff{-\frac12+\ep,-1+\ep-n,-n}{-1+2\ep,\ep}
\nonumber\\&&{}
+\frac{2(-1+2\ep)_n}{1-\ep}
\Fh32\Ff{\frac12,1-\ep-n,-n}{2-\ep,\ep},
\label{c2}\\
C_3(n)&=&
-\frac{(1-2\ep)n!}{(1-\ep)^2}\left(
\Fh32\Ff{\frac32-\ep,1-\ep-n,-n}{2-\ep,3-2\ep}
\right.\nonumber\\&&{}
+\left.\frac{2(\ep)_n}{(2-\ep)_n}
\Fh32\Ff{\frac12,-1+\ep-n,-n}{2-\ep,\ep}\right),
\label{c3}
\end{eqnarray}
for the coefficients of~(\ref{j3m}), which we have checked by substitution
in the differential equation~(\ref{jde}).

{From} the expansions of $J_3$, we generate those of
\begin{equation}
K_3\equiv\int\!\!\int\frac{\rd^dk_1\rd^dk_2}{\pi^d\Gamma^2(1+\ep)}
P(k_1,0)P(k_2,m)P(k_2-q,m)P(k_1-k_2,m),
\label{k3}
\end{equation}
using integration by parts, which yields
\begin{eqnarray}
&&(q^2+m^2)K_3-\Df13(q^2+9m^2)J_3^\prime+(3d-8)J_3
\nonumber\\&&{}
=\frac{4m^{2d-10}}{(d-3)(d-4)}
\left(\frac{m^2(q^2-m^2)}{d-4}+\frac{(q^2+m^2)^2G_2}{d-2}\right),
\label{kde}
\end{eqnarray}
with $G_2$ given by~(\ref{gh2}) at small $q^2$ and by~(\ref{f2}) at large
$q^2$. Finally we obtain $I_3$ from integration by parts,
which yields
\begin{eqnarray}
&&(d-4)((q^2+3m^2)I_3-\Df12(q^2-3m^2)I_2)
-m^2(q^2+m^2)(2I_3^\prime+I_2^\prime)+K_3-J_3^\prime
\nonumber\\&&{}
=\frac{4m^{2d-12}}{d-3}\left(\frac{2m^4}{(d-4)^2}
+\frac{m^2(q^2+m^2)G_2}{(d-2)(d-4)}-\frac{q^2(q^2+m^2)G_2^2}{(d-2)^2}\right),
\label{ide}
\end{eqnarray}
with $I_2$ given by~(\ref{i2q},\ref{i2m}). {From} this first-order equation
the expansion coefficients may be obtained, at both large and small
$q^2$, for any $d$.

The small-$q^2$ expansion is of the form
\begin{eqnarray}
-m^{4\ep+2}\ep^2(1-\ep)(1-2\ep)I_3&=&
\frac{\delta}{2}\sum_{n=0}^\infty\frac{(-q^2/2m^2)^n}{(d/2)_n n!}P_1(n)
+C(0)\sum_{n=0}^\infty\frac{(q^2/18m^2)^n}{(d/2)_n n!}P_2(n)
\nonumber\\&&{}
-6\sum_{n=0}^\infty\frac{(q^2/18m^2)^n}{(d/2)_n n!}P_3(n)
\!\!\prod_{n/2>k\geq0}\frac{1}{d+2k-2}\,,
\label{i3q}
\end{eqnarray}
where $\delta$ is given by~(\ref{del}), $C(0)$ is the constant
of~(\ref{j3z}), the final product is set to unity for $n=0$,
and $P_i(n)$ are polynomials in $d$. Table~1 gives the
first 5 terms in the Taylor series. (We have computed the first 20.)
\begin{table}[htb]
\caption{Expansion coefficients of $I_3$ at small $q^2$.}
\vspace{2.0mm}
$P_1(0)=1,\quad P_1(1)=d^2-9d+22,\quad P_1(2)=d^4-21d^3+168d^2-604d+832,$\\
$P_1(3)=d^6-36d^5+541d^4-4338d^3+19588d^2-47304d+48000,\quad
P_1(4)=d^8-54d^7$\\$\quad{}
+1271d^6-17022d^5+141884d^4-754056d^3+2497696d^2-4722432d+3916800,$\\[3pt]
$P_2(0)=1,\quad P_2(1)=d^2+7d-66,\quad P_2(2)=d^4+15d^3-28d^2-2484d+10944,$\\
$P_2(3)=d^6+24d^5+133d^4-5574d^3-62540d^2+1058280d-3248640,\quad
P_2(4)=d^8+34d^7$\\$\quad{}
+439d^6-6854d^5-278612d^4+309976d^3+69223392d^2-615536640d+1505018880,$\\[3pt]
$P_3(0)=\frac13,\quad P_3(1)=d^3-7d^2+10d-6,\quad
P_3(2)=d^5+16d^4-463d^3+2960d^2-7308d+7488$,\\
$P_3(3)=d^8+25d^7+193d^6-24269d^5+313846d^4-
1598060d^3+3568632d^2-2875392d-839808,$\\
$P_3(4)=d^{10}+35d^9+509d^8-4471d^7-1276222d^6+
28901060d^5-252836984d^4$\\$\quad{}
+1041981408d^3-1924952832d^2+880201728d+1209323520.$
\end{table}

The large-$q^2$ expansion is of the form
\begin{equation}
q^2m^{4\ep}\ep^2(1-2\ep)I_3=
\sum_{n=0}^\infty\left(\beta^2 M_0(n)-\gamma M_1(n)+\beta M_2(n)+M_3(n)\right)
\frac{(-m^2/q^2)^n}{n!},
\label{i3m}
\end{equation}
where the coefficients are given by the recurrence relations
\begin{eqnarray}
(n+\ep)M_0(n)&=&n(n-1+3\ep)M_0(n-1)+(-1+4\ep)_n,
\label{m0}\\
(n+\ep)M_1(n)&=&2n(n-1+2\ep)M_1(n-1)-n(n-1)(n-2+3\ep)M_1(n-2)
\nonumber\\&&{}
+\Df13C_1(n+1)+(n-2+3\ep)C_1(n)
\nonumber\\&&{}
+\Df12(2n-1+3\ep)(-1+2\ep)_n(3\ep)_{n-1}/(1+\ep)_n,
\label{m1}\\
n M_2(n)&=&2n(n-1+\ep)M_2(n-1)-n(n-1)(n-2+2\ep)M_2(n-2)
\nonumber\\&&{}
+\Df13(n+1-\ep)C_2(n)+n(n-2+2\ep)C_2(n-1)
\nonumber\\&&{}
+(2n-1+\ep)(-1+2\ep)_n(\ep)_{n-1}/(1-\ep)_n
\nonumber\\&&{}
-2(-2+2\ep)_n\Fh32\Fu{1,\ep,-n}{1-\ep,3-2\ep-n},
\label{m2}\\
(n-\ep)M_3(n)&=&2n(n-1)M_3(n-1)-n(n-1)(n-2+\ep)M_3(n-2)+2M_4(n)
\nonumber\\&&{}
-\Df13n(n+1-2\ep)C_3(n-1)-n(n-1)(n-2+\ep)C_3(n-2),
\label{m3}\\
(n-2\ep)M_4(n)&=&n(n-3+2\ep)(M_4(n-1)-\delta_{n,2})
\nonumber\\&&{}
-2\ep(1-2\ep)(n!-\delta_{n,0}-\Df12\delta_{n,1})(\ep)_{n-2}/(2-\ep)_{n-1},
\label{m4}
\end{eqnarray}
with $M_{0,1}(0)=1/\ep$, $M_{2,3,4}(0)=0$, and $C_{1,2,3}(n)$ given
by~(\ref{c1},\ref{c2},\ref{c3}). The extra set of coefficients
in~(\ref{m4}) serves to expand the square of series~(\ref{f2}), needed for
the $I_2$ term on the left-hand side of~(\ref{ide}) and for the $G_2^2$
term on the right. Relations~(\ref{m0}--\ref{m4}) allow very rapid
computation of the asymptotic expansion~(\ref{i3m}); the hard work has been
done in obtaining the closed forms~(\ref{c1}--\ref{c3}), for the
expansion~(\ref{j3m}) of the irreducibly difficult integral~(\ref{j3}),
together with
the differential equations~(\ref{kde},\ref{ide}). Table~2 gives the first 5
terms in the asymptotic expansion. (We have computed the first 20.)
\begin{table}[htb]
\caption{Expansion coefficients of $I_3$ at large $q^2$.}
\vspace{2.0mm}
$M_0(0)={-2\over (d-4)},\quad
 M_0(1)={4(d-5)\over (d-6)},\quad
 M_0(2)={-4(2d^3-33d^2+176d-308)\over (d-6)(d-8)},$\\
$M_0(3)={8(2d^3-37d^2+210d-384)(d-5)(d-7)\over (d-6)(d-8)(d-10)},$\\
$M_0(4)={-8(4d^5-144d^4+1987d^3-13158d^2+42064d-52416)(d-5)
 \over (d-8)(d-10)(d-12)},$\\[3pt]
$M_1(0)={-2\over (d-4)},\quad
 M_1(1)={12(d-4)(d-5)\over (d-6)^2},\quad
 M_1(2)={-6(5d^3-86d^2+472d-800)(3d-14)(d-5)\over (d-6)^2(d-8)^2},$\\
$M_1(3)={12(7d^4-195d^3+1952d^2-8172d+11712)(3d-14)(3d-16)(d-5)(d-7)
 \over (d-6)^2(d-8)^2(d-10)^2},$\\
$M_1(4)={-18(41d^6-2046d^5+41532d^4-436680d^3+2489536d^2-7211904d+8128512)
 (3d-14)(3d-16)(d-5)(d-7)\over (d-6)(d-8)^2(d-10)^2(d-12)^2},$\\[3pt]
$M_2(0)=0,\quad
 M_2(1)={2(d-3)(d-8)\over (d-2)},\quad
 M_2(2)={-2(d^3-15d^2+22d+152)(d-3)\over (d-2)d},$\\
$M_2(3)={4(2d^5-45d^4+197d^3+108d^2+3500d-23040)(d-3)
 \over 3(d+2)(d-2)d},$\\
$M_2(4)={-4(2d^7-51d^6+90d^5+3063d^4-6180d^3+42420d^2-1300496d+4660992)
 (d-3)\over 3(d+4)(d+2)(d-2)d},$\\[3pt]
$M_3(0)=0,\quad
 M_3(1)={4\over (d-2)},\quad
 M_3(2)={-8(d^2-19d+44)\over (d-2)^2d},$\\
$M_3(3)={-24(39d^2-482d+1072)\over (d+2)(d-2)^2d},\quad
 M_3(4)={192(d^5+21d^4-824d^3+7104d^2-11672d-8352)
 \over (d+4)(d+2)(d-2)^2d^2}.$
\end{table}

Taking the limit $d\to4$, we obtain, from Tables~1 and~2,
the expansions given in~\cite{DJB} for
\begin{eqnarray}
&&\left.q^2I_3\right|_{d=4}=\int_{m}^\infty\frac{4w\,\rd w}{w^2-m^2}
\left(\ln\frac{w}{m}-\frac{w^2-m^2}{w^2}\ln\frac{w^2-m^2}{m^2}\right)
\ln\frac{w^2+q^2}{w^2}
\nonumber\\&&{}
+\int_{2m}^{\infty}\frac{4\,\rd w}{(w^2-4m^2)^{1/2}}
\left(3\ln\frac{w}{m}
-\frac{3w^2-3m^2+q^2}{W_+W_-}\ln\frac{W_++W_-}{W_+-W_-}
\right)\ln\frac{w}{2m},
\label{i34}
\end{eqnarray}
with $W_\pm\equiv((w\pm m)^2+q^2)^{1/2}$.
(Note that the sign of $q^2$
in~\cite{DJB} is opposite to that chosen here.)
In the case $d=3$, we find that $I_3$ is given by the imaginary part of a
complex dilogarithm, throughout the space-like region $q^2>0$, and is hence
reducible to three instances of Clausen's integral~\cite{LEW}:
\begin{equation}
\left.m^4I_3\right|_{d=3}
=\frac{2\,{\rm Im}\Li2\left(r{\rm e}^{\ri \theta}\right)}
{r\tan(\theta/2)}
=\frac{2\omega\ln r
+{\rm Cl}_2(2\omega)
+{\rm Cl}_2(2\theta)
-{\rm Cl}_2(2\omega+2\theta)}
{r\tan(\theta/2)}\,,
\label{i33}
\end{equation}
with
\begin{equation}
r=\frac{m^2+q^2}{4m^2}\,,\quad
\cos\theta=\frac{m^2-q^2}{m^2+q^2}\,,\quad
\cos\omega=\frac{3m^2+q^2}{(9m^2+q^2)^{1/2}(m^2+q^2)^{1/2}}\cdot
\label{clau}
\end{equation}

\section{The problem of intermediate $q^2$}

In the cases of integrals $I_1$, $I_4$, $I_2$ and $\widetilde{I}_3$, the
small-$q^2$ expansions~(\ref{i1q},\ref{i4q},\ref{i2q},\ref{i3tq}) and the
large-$q^2$ expansions~(\ref{i1m},\ref{i4m},\ref{i2m},\ref{i3tm}) together
cover the entire complex $q^2$-plane, cut along the negative real axis,
with the possible exception of the circle $|q^2|=m^2$, in the cases of
$I_1$ and $I_2$, or the circle $|q^2|=4m^2$, in the cases of $I_4$ and
$\widetilde{I}_3$. On these circles, the series may converge only
conditionally, or may diverge, depending on the value of $d$. Otherwise,
one has a method for computing the integrals by truncation of the
appropriate series, for any $d$ and $q^2$.

This happy circumstance is unfortunately far from generic. In the general
mass case, the integral
\begin{equation}
I_5=\int\!\!\int\frac{\rd^dk_1\rd^dk_2}{\pi^d\Gamma^2(1+\ep)}
P(k_1,m_1)P(k_2,m_2)P(k_1-q,m_3)P(k_2-q,m_4)P(k_1-k_2,m_5),
\label{i5}
\end{equation}
has four contributions, with distinct branchpoints, corresponding to the
four ways of cutting the master diagram~\cite{DJB}. The small-$q^2$ expansion
converges for
\begin{equation}
|q^2|<{\rm Min}\left((m_1+m_3)^2,\,(m_2+m_4)^2,\,
(m_1+m_4+m_5)^2,\,(m_2+m_3+m_5)^2\right),
\label{qmin}
\end{equation}
whilst the large-$q^2$ expansion converges for
\begin{equation}
|q^2|>{\rm Max}\left((m_1+m_3)^2,\,(m_2+m_4)^2,\,
(m_1+m_4+m_5)^2,\,(m_2+m_3+m_5)^2\right).
\label{qmax}
\end{equation}
In the case of a branchpoint at the origin, as occurs for $I_1$ and
$\widetilde{I}_3$, a small-$q^2$ expansion may be found, valid up to the
next branchpoint, as in~(\ref{i1q},\ref{i3tq}). In the general mass case,
however, each method fails at intermediate values of $|q^2|$. In
particular, $I_3$ cannot be computed directly from the results of Tables~1
and~2 when $q^2$ is in the problematic annulus $9m^2>|q^2|>m^2$.

Tables~3 and~4 show the accuracies to which truncations of the small and
large $q^2$ series approximate the exact results~(\ref{i34},\ref{i33}), and
their analytic continuations to the time-like region $-q^2\in[0,m^2]$.
They show the relative error, $\Delta_d(n)\equiv|I_3^{\rm trunc}/I_3-1|$,
when $I_3^{\rm trunc}$ is obtained from the first $n$ terms in the series,
with $d=4$ or $d=3$.

The domain of convergence of the small-$q^2$ Taylor series of Table~3 is
$|q^2|\leq m^2$, for $d=4$, and $|q^2|<m^2$, for $d=3$. There is a
branchpoint at $q^2=-m^2$, where a cut begins, due to intermediate states
with only one massive particle. The limit of~(\ref{i34}) exists
as $q^2\to-m^2$, giving~\cite{DJB} $m^2I_3\to6\zeta(2)\ln2-\Df32\zeta(3)$,
for $d=4$. The limit of~(\ref{i33}) does not exist; instead one finds that
$m^2(m^2+q^2)I_3\to4\zeta(2)$, for $d=3$. Note that the convergence of the
$d=4$ Taylor series is slow for $|q^2|=m^2$.

One expects the asymptotic truncation errors to become substantial for
$q^2\approx9m^2$, since the intermediate states with three massive
particles generate a branchpoint at $q^2=-9m^2$. The last row of Table~4
shows that one may trespass only a small distance inside the {\em annulus
horribilis} before the divergence of the asymptotic expansion becomes a
problem. The task of computing the integral reliably, by series expansions
at intermediate $q^2$, is clearly a delicate one, which we accomplish in
Section~3.1, by separating the sources of discontinuity. In Section~3.2, we
show how to avoid this separation, for space-like $q^2$, by using methods
of accelerated convergence~\cite{EPS}.
\begin{table}[htb]
\caption{Small-$q^2$ truncation errors.}
\[\begin{array}{|c|lll|lll|}\hline q^2/m^2&
{}~~\Delta_4(5) & ~~\Delta_4(10)& ~~\Delta_4(15)&
{}~~\Delta_3(5) & ~~\Delta_3(10)& ~~\Delta_3(15)                \\ \hline
\Pm 1.0 & \Pp672 & \Pp232 & \Pp122 &        &        &        \\
\Pm 0.8 & \Pp232 & \Pp263 & \Pp444 & \Pp511 & \Pp171 & \Pp572 \\
\Pm 0.6 & \Pp563 & \Pp154 & \Pp616 & \Pp121 & \Pp933 & \Pp734 \\
\Pm 0.4 & \Pp774 & \Pp286 & \Pp158 & \Pp152 & \Pp164 & \Pp166 \\
\Pm 0.2 & \Pp255 & \Pp299 &\Pp48{13}&\Pp454 & \Pp157&\Pp48{11}\\
   -0.2 & \Pp295 & \Pp349 &\Pp58{13}&\Pp424 & \Pp147&\Pp44{11}\\
   -0.4 & \Pp103 & \Pp406 & \Pp228 & \Pp132 & \Pp134 & \Pp146 \\
   -0.6 & \Pp883 & \Pp274 & \Pp125 & \Pp912 & \Pp723 & \Pp574 \\
   -0.8 & \Pp462 & \Pp643 & \Pp123 & \Pp361 & \Pp121 & \Pp392 \\
   -1.0 & \Pp271 & \Pp171 & \Pp121 &        &        &        \\ \hline
\end{array}\]
\end{table}
\begin{table}[htb]
\caption{Large-$q^2$ truncation errors.}
\[\begin{array}{|c|lll|lll|}\hline q^2/m^2&
{}~~\Delta_4(5) & ~~\Delta_4(10)& ~~\Delta_4(15)&
{}~~\Delta_3(5) & ~~\Delta_3(10)& ~~\Delta_3(15)             \\ \hline
  15 & \Pp185 & \Pp257 & \Pp649 & \Pp183 & \Pp335 & \Pp116 \\
  13 & \Pp485 & \Pp136 & \Pp678 & \Pp383 & \Pp144 & \Pp105 \\
  11 & \Pp154 & \Pp866 & \Pp106 & \Pp913 & \Pp804 & \Pp134 \\
\Pu9 & \Pp534 & \Pp815 & \Pp255 & \Pp262 & \Pp623 & \Pp273 \\
\Pu7 & \Pp263 & \Pp133 & \Pp143 & \Pp992 & \Pp812 & \Pp121 \\ \hline
\end{array}\]
\end{table}
\subsection{Separating the discontinuities}

Denoting~\cite{DJB} the two contributions to~(\ref{i34}) as $I_a$ and
$I_b$, we find separate asymptotic expansions of the form
\begin{equation}
I_{a,b}=3\zeta(3)\mp\left(\Df16L^3-2\zeta(2)L+7\zeta(3)\right)
+\sum_{n=1}^\infty\left(\frac{a_{a,b}(n)}{n}L^2+\frac{b_{a,b}(n)}{n^2}L
+\frac{c_{a,b}(n)}{n^3}\right)\left(\frac{-m^2}{q^2}\right)^n,
\label{iabm}
\end{equation}
with $L\equiv\ln(q^2/m^2)$. The coefficients of the simpler term,
$I_a$, with no contribution from intermediate states with three massive
particles, may be obtained in closed form:
\begin{equation}
a_a(n)=-\Df12,\quad b_a(n)=-3,
\quad c_a(n)=-4-2n^2\zeta(2)+\sum_{r=1}^{n-1}\frac{n}{r}
\left(2-\frac{n}{r}\right),
\label{iam}
\end{equation}
from which we obtain the coefficients of $I_b$ by subtraction, given the
$d\to4$ limit of the asymptotic expansion~(\ref{i3m}).
The small-$q^2$ expansions are of the form
\begin{equation}
I_{a,b}=-\sum_{n=1}^\infty\frac{d_{a,b}(n)}{n}\left(\frac{-q^2}{m^2}\right)^n,
\quad d_a(n)=\zeta(2)+\sum_{r=1}^{n-1}\frac{1}{rn}\left(2-\frac{n}{r}\right),
\label{iabq}
\end{equation}
from which we obtain the coefficients of $I_b$ from the $d\to4$
limit of the Taylor series~(\ref{i3q}).

The strategy now is clear: in the annulus one uses the large-$q^2$
expansion of $I_a$ and the small-$q^2$ expansion of $I_b$. (The small-$q^2$
expansion of $I_a$ and the large-$q^2$ expansion of $I_b$ are never needed;
outside the annulus one deals with the full integral.) A computational
difficulty becomes apparent, however, when one evaluates the small-$q^2$
coefficients of the troublesome term, $I_b$: the coefficient $d_b(n)$ is
numerically undetermined unless the constant~(\ref{s2}) is known to an
accuracy substantially better than 1 part in $9^n$. For example,
\begin{equation}
d_b(20)=-2\zeta(2)
-\Df{57479895945135907790704519109851}{14345652399118288857618205996800}
+\Df{934497714974357881}{33354363399333138}S_2
\approx6.074\times10^{-22}
\label{d20}
\end{equation}
is 22 orders of magnitude smaller than its $S_2$-dependent term. Yet, at
$q^2=9m^2$, its relative contribution to the Taylor series is
$2\times10^{-4}$.

It may now be seen that the method of~\cite{INLO}, which gives the Taylor
series of Feynman integrals in terms of $S_2$, requires one to work with
intermediate numerical precision much greater than the desired final
accuracy: 18 orders of magnitude greater in the above example. Fortunately
one may exploit this novel feature and use the near vanishing of higher
coefficients in the series to obtain a suitably accurate approximation to
$S_2$, which is how we computed the value~(\ref{s2v}), using the analytical
expression for $d_b(30)$. By high precision computation of the appropriate
series, we then obtained the acceptably small truncation errors of Table~5.
\begin{table}[htb]
\caption{Intermediate-$q^2$ truncation errors.}
\[\begin{array}{|c|lll|lll|}\hline q^2/m^2&
{}~~\Delta_4(5) & ~~\Delta_4(10)& ~~\Delta_4(15)&
{}~~\Delta_3(5) & ~~\Delta_3(10)& ~~\Delta_3(15)           \\ \hline
8  & \Pp103 & \Pp875 & \Pp155 & \Pp763 & \Pp133 & \Pp344 \\
7  & \Pp494 & \Pp225 & \Pp206 & \Pp363 & \Pp304 & \Pp415 \\
6  & \Pp224 & \Pp456 & \Pp197 & \Pp203 & \Pp575 & \Pp366 \\
5  & \Pp995 & \Pp707 & \Pp128 & \Pp213 & \Pp846 & \Pp207 \\
4  & \Pp885 & \Pp107 &\Pp42{10}&\Pp593 & \Pp696 & \Pp128 \\
3  & \Pp364 & \Pp787 & \Pp229 & \Pp302 & \Pp144 & \Pp597 \\
2  & \Pp403 & \Pp695 & \Pp156 & \Pp321 & \Pp112 & \Pp364 \\ \hline
\end{array}\]
\end{table}

For $d=3$, one must separate~(\ref{i33}) as follows:
\begin{equation}
\left.m^4I_3\right|_{d=3}=G_a+G_b,\quad
G_a=\frac{\theta\ln r+4{\rm Cl}_2(\pi-\theta)-\zeta(2)\tan(\theta/2)}
{r\tan(\theta/2)}
\label{i3ab}
\end{equation}
and use the asymptotic expansion of $G_a$ and the Taylor series of $G_b$
in the annulus. For the latter, high precision is again required, to deal
with cancellations of the type
\begin{equation}
\zeta(2)
+\Df{9291381626641020913518612265142}{3909280494259534670432360629425}
-\Df{1656090499861696}{166966608033225}\ln\Df32 \approx
1.076\times10^{-22},
\label{das}
\end{equation}
which occurs in the coefficient analogous to~(\ref{d20}).

\subsection{Accelerated convergence}

Another strategy for dealing with the problematic region $q^2\in[m^2,9m^2]$
is the use of techniques that accelerate the convergence of the large and
small $q^2$ expansions. A convenient technique is the $\ep$-algorithm
of~\cite{EPS}. In general, given a sequence of approximations,
$\{S_n|n=0,1,2,\ldots\}$, one constructs a table of approximants using:
\begin{equation}
T(m,n)=T(m-2,n+1)+1/\left\{T(m-1,n+1)-T(m-1,n)\right\},
\label{eps}
\end{equation}
with $T(0,n)\equiv S_n$ and $T(-1,n)\equiv 0$. If the sequence $\{S_n\}$ is
obtained by successive truncation of a Taylor series, the approximant
$T(2k,j)$ is identical to the $[k+j/k]$ Pad\'{e} approximant~\cite{EPS},
derived from the first $2k+j+1$ terms in the Taylor series. Moreover, in
the case of Stieltjes series, with a constant-sign spectral density, the
$n\to\infty$ limit of diagonal Pad\'{e} approximants converges throughout
the space-like region, giving the analytic continuation of the function
whose derivatives at the origin are the Taylor coefficients. We have
verified that~(\ref{i34},\ref{i33}) are of Stieltjes type.
\begin{table}[htb]
\caption{Accelerated convergence of the Taylor series of $I_3$,
with $d=4$.}
\[\begin{array}{|c|llll|}\hline
q^2/m^2 &~~E_4(2)   &~~E_4(4)   &~~E_4(6)   &~~E_4(8)   \\ \hline
\Pm 3.0 & \Pp603    & \Pp555    & \Pp587    & \Pp599    \\
\Pm 2.0 & \Pp193    & \Pp756    & \Pp348    & \Pp15{10} \\
\Pm 1.0 & \Pp194    & \Pp137    & \Pp10{10} & \Pp82{14} \\
\Pm 0.8 & \Pp835    & \Pp308    & \Pp13{11} & \Pp52{15} \\
\Pm 0.6 & \Pp275    & \Pp419    & \Pp70{13} & \Pp12{16} \\
\Pm 0.4 & \Pp506    & \Pp20{10} & \Pp93{15} & \Pp43{19} \\
\Pm 0.2 & \Pp237    & \Pp82{13} & \Pp32{18} & \Pp13{23} \\
   -0.2 & \Pp577    & \Pp47{12} & \Pp42{17} & \Pp39{22} \\
   -0.4 & \Pp325    & \Pp739    & \Pp18{12} & \Pp44{16} \\
   -0.6 & \Pp514    & \Pp116    & \Pp289    & \Pp65{12} \\
   -0.8 & \Pp573    & \Pp114    & \Pp226    & \Pp388    \\
   -1.0 & \Pp121    & \Pp462    & \Pp182    & \Pp433    \\ \hline
\end{array}\]
\end{table}

Table~6 shows the errors $E_4(n)\equiv|I_3^\ep/I_3-1|$, with $I_3^\ep$
obtained, via the $\ep$-algorithm, from $T(2n,0)$, i.e.\ from the $[n/n]$
Pad\'{e} approximant to the first $2n+1$ terms of the Taylor series of
$I_3$, with $d=4$. Comparison of Tables~3 and~6 reveals the accelerated
convergence for $|q^2|\leq m^2$. Table~6 also shows the extension of the
domain of convergence to $q^2>m^2$.

For larger values of $q^2$, we apply the $\ep$-algorithm to the asymptotic
expansion of $I_3$. Now the $n$th term in the series is $(-m^2/q^2)^n$
times a quadratic polynomial in $\ln(q^2/m^2)$, and hence there is no
Pad\'{e} approximant corresponding to the first $2n+1$ terms, nor any
theorem, known to us, to guarantee convergence of the $\ep$-algorithm
for $|q^2|<9m^2$, outside the domain of convergence of the
asymptotic expansion. Nevertheless, remarkably good results are obtained
in the space-like region $q^2\geq3m^2$, as can be seen from Table~7.

Comparison of Tables~4 and~7 reveals accelerated convergence, in an
enlarged domain, as was observed at small $q^2$. Together, Tables~6 and~7
show that results accurate to a few parts in $10^9$ can be obtained {\em
throughout\/} the space-like region, $q^2>0$. Thus, if one needs only
space-like values, the delicate separation method of Section~3.1 may be
avoided; it suffices to use the $\ep$-algorithm, to achieve accelerated
convergence of the large-$q^2$ expansion, for $q^2\geq3m^2$, or the
small-$q^2$ expansion, for $q^2\leq3m^2$. Only if one requires the real
and imaginary parts on the problematic region of the cut, with $-q^2=s+{\rm
i}0$ and $s\in[m^2,9m^2]$, does one need to disentangle
the separate sources of discontinuity.
\begin{table}[htb]
\caption{Accelerated convergence of the asymptotic expansion of $I_3$,
with $d=4$.}
\[\begin{array}{|c|llll|}\hline
q^2/m^2 &~~E_4(2)   &~~E_4(4)   &~~E_4(6)   &~~E_4(8)   \\ \hline
\Pm12.0 & \Pp897    & \Pp229    & \Pp57{14} & \Pp90{17} \\
\Pm11.0 & \Pp206    & \Pp769    & \Pp86{14} & \Pp57{16} \\
\Pm10.0 & \Pp426    & \Pp458    & \Pp80{13} & \Pp60{15} \\
\Pm 9.0 & \Pp906    & \Pp147    & \Pp41{12} & \Pp36{14} \\
\Pm 8.0 & \Pp205    & \Pp978    & \Pp20{11} & \Pp46{14} \\
\Pm 7.0 & \Pp465    & \Pp147    & \Pp11{10} & \Pp17{13} \\
\Pm 6.0 & \Pp124    & \Pp277    & \Pp67{10} & \Pp42{15} \\
\Pm 5.0 & \Pp334    & \Pp617    & \Pp539    & \Pp14{12} \\
\Pm 4.0 & \Pp113    & \Pp146    & \Pp618    & \Pp46{11} \\
\Pm 3.0 & \Pp533    & \Pp186    & \Pp176    & \Pp179    \\ \hline
\end{array}\]
\end{table}
\section{Photon self-energy}

Evaluating two-loop diagrams for the photon self-energy,
one encounters the integrals
\begin{eqnarray}
&&\int\!\!\int\frac{\rd^dk_1\rd^dk_2}{\pi^d\Gamma^2(1+\ep)}
P^{j_1}(k_1,m)P^{j_2}(k_2,m)P^{j_3}(k_1-q,m)P^{j_4}(k_2-q,m)
P^{j_5}(k_1-k_2,0)
\nonumber\\&&{}
\quad\quad\quad\equiv V_Q(j_1,j_2,j_3,j_4,j_5)
(m^2)^{d-j_1-j_2-j_3-j_4-j_5},
\label{VQ}
\end{eqnarray}
with the propagators of Fig.~1(d) raised to integer powers $j_1$ to $j_5$.
Integration by parts yields the recurrence relations
\begin{eqnarray}
T(j_5,j_2,j_4)V_Q&=&\left\{
j_2{\bf 2}^{+}({\bf 5}^{-}-{\bf 1}^{-})+
j_4{\bf 4}^{+}({\bf 5}^{-}-{\bf 3}^{-})\right\}V_Q\,,
\label{kill1}\\
T(j_5,j_1,j_3)V_Q&=&\left\{
j_1{\bf 1}^{+}({\bf 5}^{-}-{\bf 2}^{-})+
j_3{\bf 3}^{+}({\bf 5}^{-}-{\bf 4}^{-})\right\}V_Q\,,
\label{kill2}
\end{eqnarray}
where $T(j_k,j_l,j_m)\equiv d-2j_k-j_l-j_m$ and ${\bf
1}^{\pm}V_Q(j_1,...)\equiv V_Q(j_1\pm1,...)$, etc. Using~(\ref{kill1}) to
remove one propagator, carrying the momentum $k_1$, and then~(\ref{kill2})
to remove another, carrying the momentum $k_2$, every $V_Q$ integral can be
reduced to combinations of a rational function of $q^2/m^2$ and $d$
(deriving from tadpole diagrams, through which $q$ does not flow) and
rational multiples of integrals of the type
\begin{eqnarray}
V_Q(2,2,1,0,0)&=&\frac{G_4}{4-d}\,,
\label{gg4}\\
V_Q(2,2,1,1,0)&=&\Df14G_4^2\,,
\label{gg42}\\
V_Q(0,\alpha,\beta,0,\gamma)&=&\frac{
\Gamma(\alpha+\beta+\gamma-d)\Gamma(\frac{d}{2}-\gamma)
\Gamma(\beta+\gamma-\frac{d}{2})\Gamma(\alpha+\gamma-\frac{d}{2})}
{\Gamma(\alpha)\Gamma(\beta)\Gamma(\frac{d}{2})
\Gamma(\alpha+\beta+2\gamma-d)\Gamma^2(3-\frac{d}{2})}
\nonumber\\&&
\cdot\Fh43\Ffq{\gamma,\alpha+\beta+\gamma-d,\beta+\gamma-
\frac{d}{2},\alpha+\gamma-\frac{d}{2}}
{\frac{d}{2},\gamma+\frac{\alpha+
\beta-d}{2},\gamma+\frac{\alpha+\beta-d+1}{2}}\,.
\label{gabc}
\end{eqnarray}
Moreover, the $\Fh43$ hypergeometric functions of~(\ref{gabc}), with
integer values of $\alpha$, $\beta$ and $\gamma$, are always reducible to
two contiguous $\Fh32$ functions.

It follows that {\em every\/} two-loop calculation of the
perturbative~\cite{PLB,SCG} contribution, or of a
gluon-condensate~\cite{GLUE} contribution, to the correlator of a current
connecting quarks of equal mass, amounts to no more than the evaluation of
5 rational functions, corresponding to the coefficients of unity, $G_4$,
$G_4^2$, and two suitably chosen $\Fh32$ functions. In the case of the
master integral $I_4$, with unit exponents, the functions $G_4$ and $H_4$
were chosen in~(\ref{gh4}) in such a way that only $G_4^2$ and $H_4$
appeared in the result~(\ref{i4q}). In general, a basis
$\{1,G_4,G_4^2,I_4,I_4^\prime\}$ gives coefficients that are regular as
$d\to4$, enabling one to obtain as many terms as desired in the small-$q^2$
expansion, or the large-$q^2$ expansion, merely by taking the $d\to4$
limits of the 5 rational coefficients, in this basis, and using the $d=4$
expansions of $G_4$ and $I_4$. In this sense, $I_4$ is truly the {\em
master} integral, for this mass case.

\subsection{Vacuum polarization function}

In the on-shell renormalization scheme of conventional QED~\cite{KS,JS,BR},
the renormalized photon propagator has a denominator $(1+\Pi)$, where
the vacuum polarization function, $\Pi$, vanishes at $q^2=0$.
Generalizing to arbitrary $d$, we write
\begin{equation}
\Pi(z)=\frac{T(d)}{4}\Bigg\{
\left(\frac{\alpha}{4\pi}\right)\Pi_1(z)+
\left(\frac{\alpha}{4\pi}\right)^2\Pi_2(z)+O(\alpha^3)\Bigg\},
\label{Piz}
\end{equation}
where $z\equiv-q^2/4m^2$, traces are normalized by ${\rm
Tr}(\gamma_\mu\gamma_\nu)=T(d)g_{\mu\nu}$, and we form the dimensionless
coupling $\alpha\equiv\Gamma(3-d/2)m^{d-4}e^2/(4\pi)^{d/2-1}$ from the
on-shell charge and mass. The renormalization-scheme-invariant coupling is
given by $\alpha/(1+\Pi)$. For $d=4$, $\alpha=e^2/4\pi$ is the fine
structure constant of QED, measured at $q^2=0$.

To give a compact form for $d$\/-dimensional two-loop vacuum
polarization, we choose the basis $\{1,\phi_1,\phi_1^2,\phi_2,\phi_3\}$,
where
\begin{eqnarray}
\phi_1(z)&=&z\,\frac{(d-6)}{5}\Fh21\Ffz{1,4-\frac{d}{2}}{\frac72},
\label{phi1}\\
\phi_2(z)&=&\frac{(d-3)(d-4)(d-6)}{(d+2)d(d-5)(d-7)}
\Fh32\Ffz{1,4-\frac{d}{2},5-d}{2+\frac{d}{2},\frac{9-d}{2}},
\label{phi2}\\
\phi_3(z)&=&\frac{(d-6)}{(d+2)d(d-1)(d-7)}
\Fh32\Ffz{1,4-\frac{d}{2},6-d}{2+\frac{d}{2},\frac{9-d}{2}}.
\label{phi3}
\end{eqnarray}
We implemented~(\ref{kill1}--\ref{gabc}) in both~\cite{RED} {\sc Reduce},
running on a VAX4100, and {\sc Form}, running on a AT486. In each case,
a few seconds of CPUtime yielded
\begin{eqnarray}
&&\!\!\!\Pi_1(z)=
\frac{4}{3(d-1)}\left\{z(d-2)-\left[1+z(d-2)\right]\phi_1(z)\right\},
\label{Pi1}\\
&&\!\!\!\Pi_2(z)=\frac{-32z}{9(d-4)}\Bigg\{
\frac{d^4-12d^3+48d^2-53d-24}{d(d-1)(d-5)}
-\frac{z(d-2)(d^2-7d+16)\left[1-\phi_1(z)\right]^2}{4(d-1)}
\nonumber\\&&\quad{}
+\frac{\left[d-6-(d-3)\phi_1(z)\right]^2}{2(d-3)(1-z)}
-\frac{(d-2)(d-5)(d-6)}{2(d-1)(d-3)}\,\phi_1(z)
-\frac{(d-2)^2}{2(d-1)}\,\phi_1^2(z)
\nonumber\\&&\quad{}
-\frac{9}{d-3}\,\phi_2(z)
+\frac{18\left[(d-2)(d-5)-z(d^2-5d+8)\right]}{d-3}\,\phi_3(z)\Bigg\}.
\label{Pi2}
\end{eqnarray}

The $d\to4$ limit of $\Pi_2(z)$ is best taken after transforming to the
basis $\{1,G_4,G_4^2,I_4,I_4^\prime\}$, in which the coefficients
are regular, though rather lengthy (on account of factors of $(3d-8)$ and
$(3d-10)$ in their denominators). At $d=4$, with $T(4)=4$, we obtain
\begin{eqnarray}
\left.\Pi_1(z)\right|_{d=4}&=&\frac{20}{9}+\frac{4}{3z}
-\frac{4(1-z)(1+2z)}{3z}\,G(z)\,,
\label{Pi14}\\
\left.\Pi_2(z)\right|_{d=4}&=&\frac{10}{3}+\frac{26}{3z}
-\frac{4(1-z)(3+2z)}{z}\,G(z)
+\frac{2(1-z)(1-16z)}{3z}\,G^2(z)
\nonumber\\&&{}
-\frac{2(1+2z)}{3z}\left[1+2z(1-z)\frac{{\rm d}}{{\rm d}z}\right]
\frac{I(z)}{z}\,,
\label{Pi24}\\
G(z)&\equiv&\left.G_4\right|_{d=4}=
\Fh21\Ffz{1,1}{\frac32}=\frac{2y\ln y}{y^2-1}\,,
\label{phi}\\
I(z)&\equiv&\left.q^2I_4\right|_{d=4}=
6\left[\zeta(3)+4\Li3(-y)+2\Li3(y)\right]
-8\left[2\Li2(-y)+\Li2(y)\right]\ln y
\nonumber\\&&{}\phantom{\left.q^2I_4\right|_{d=4}=}{}
-2\left[2\ln(1+y)+\ln(1-y)\right]\ln^2y\,,
\label{I4z}\\
y&\equiv&\frac{\sqrt{1-1/z}-1}{\sqrt{1-1/z}+1}\,,
\label{ydef}
\end{eqnarray}
with polylogarithms~\cite{LEW} in~(\ref{I4z}) that were obtained
in~\cite{PLB} and later derived dispersively in~\cite{DJB}. In the
next section we derive the discontinuities of~(\ref{Pi24},\ref{I4z})
from our hypergeometric functions~(\ref{phi1}--\ref{phi3}).

For $d=3$, $\Pi_2(z)$ has an infrared divergence, attributable to on-shell
singularities in the fermion propagator, which prohibit on-shell mass
renormalization. A finite result may, however, be obtained in terms of the
bare mass.

For $d=2$, with $T(2)=2$, we obtain
\begin{eqnarray}
\left.\Pi_1(z)\right|_{d=2}&=&-\frac43-\frac2{z}+\frac2{z}\,G(z)\,,
\label{Pi12}\\
\left.\Pi_2(z)\right|_{d=2}&=&
\frac{16}{3}+\frac{2(1-2z)}{z(1-z)}\,G(z)-\frac{2}{z(1-z)}\,G^2(z)\,.
\label{Pi22}
\end{eqnarray}

\subsection{Spectral density}

For the vacuum polarization function $\Pi(z)$,
we write the spectral representation
\begin{equation}
\Pi(z)=z\int_{1}^{\infty}\frac{\rho(t){\rm d}t}{t-z},
\label{disp}
\end{equation}
with spectral density
\begin{equation}
\rho(t)=\frac{{\rm Im}\,\Pi(t+\ri0)}{\pi t}=\frac{T(d)}{4}\Bigg\{
\left(\frac{\alpha}{4\pi}\right)\rho_1(t)+
\left(\frac{\alpha}{4\pi}\right)^2\rho_2(t)+O(\alpha^3)\Bigg\}.
\label{rho}
\end{equation}
{From} the general integral representations
\begin{eqnarray}
\Fh21\Ffz{a,b}{c}&=&\frac{\Gamma\left(c\right)}
{\Gamma\left(b\right)\Gamma\left(c-b\right)}
\int_{1}^{\infty}\frac{(t-1)^{c-b-1}t^{a-c}{\rm d}t}{(t-z)^a},
\label{int21}\\
\Fh32\Ffz{a,b,c}{~d,e}&=&\frac{\Gamma\left(d\right)\Gamma\left(e\right)}
{\Gamma\left(b\right)\Gamma\left(c\right)\Gamma\left(d+e-b-c\right)}
\nonumber\\&&
\cdot\int_{1}^{\infty}\frac{(t-1)^{d+e-b-c-1}t^{a-d}{\rm d}t}{(t-z)^a}
\Fh21\Fft{e-c,e-b}{d+e-b-c},
\end{eqnarray}
we obtain the spectral representations of the hypergeometric functions
in~(\ref{phi1},\ref{phi2},\ref{phi3}), which all have $a=1$.
To obtain the contribution to the
spectral density of the $\phi_1^2(z)$ terms in~(\ref{Pi2}), we use
the transformation
\begin{eqnarray}
\Fh21\Ffz{a,b}{c}&=&\frac{\Gamma\left(c\right)\Gamma\left(c-a-b\right)}
{\Gamma\left(c-a\right)\Gamma\left(c-b\right)}\Fh21\Fzz{a,b}{1+a+b-c}
\nonumber\\&&\!\!\!\!\!{}
+z^{1-c}(1-z)^{c-a-b}\frac{\Gamma\left(c\right)\Gamma\left(a+b-c\right)}
{\Gamma\left(a\right)\Gamma\left(b\right)}\Fh21\Fzz{1-a,1-b}{1-a-b+c},
\label{fff}
\end{eqnarray}
which enables one to split $\phi_1(t+\ri0)$ into its real and imaginary
parts, for $t\in[1,\infty]$. In this particular case, with $a=1$, the
second hypergeometric function in~(\ref{fff}) is trivial.

Our results for the $d$\/-dimensional spectral density are as
follows:
\begin{eqnarray}
\rho_1(t)&=&\frac{1+t(d-2)}{2t^2}\,r_1(t)\,,
\label{rho1}\\
\rho_2(t)&=&\frac{4}{(d-3)(d-4)t^2}
\Bigg\{2t^2\rho_1(t)-r_2(t)-\frac{2(d-2)(d-5)-2t(d^2-5d+8)}{d-1}\,r_3(t)
\nonumber\\&&{}
-\frac{2+t(d-2)(d-4)(d-5)-t^2(d-2)(d^2-7d+16)}{2(d-5)}\,r_1(t)\,r_4(t)
\Bigg\},
\label{rho2}\\
r_1(t)&=&\frac{\Gamma\left(\frac12\right)t^{-\frac12}(t-1)^{\frac{d-3}{2}}}
{\Gamma\left(\frac{d+1}{2}\right)\Gamma\left(3-\frac{d}{2}\right)},
\label{r1}\\
r_2(t)&=&\frac{\Gamma\left(\frac{d}{2}\right)\Gamma\left(\frac{5-d}{2}\right)
t^{1-\frac{d}{2}}(t-1)^{\frac{3d-7}{2}}}
{\Gamma\left(3-d\right)
\Gamma\left(\frac{3d-5}{2}\right)\Gamma\left(3-\frac{d}{2}\right)}
\Fh21\Fft{\frac{d-1}{2},\frac12}{\frac{3d-5}{2}},
\label{r2}\\
r_3(t)&=&\frac{\Gamma\left(\frac{d}{2}\right)\Gamma\left(\frac{5-d}{2}\right)
t^{1-\frac{d}{2}}(t-1)^{\frac{3d-9}{2}}}
{\Gamma\left(5-d\right)
\Gamma\left(\frac{3d-7}{2}\right)\Gamma\left(3-\frac{d}{2}\right)}
\Fh21\Fft{\frac{d-3}{2},\frac12}{\frac{3d-7}{2}},
\label{r3}\\
r_4(t)&=&\frac{\Gamma\left(\frac{d+1}{2}\right)
\Gamma\left(\frac{7-d}{2}\right)\sin\left(\frac{\pi d}{2}\right)}
{1-t}\,r_1(t)+\Fh21\Fft{1,3-\frac{d}{2}}
{\frac{7-d}{2}},
\label{r4}
\end{eqnarray}
where $r_{2,3}(t)$ derive from the imaginary parts of~(\ref{phi2},\ref{phi3}),
whilst $r_{4,1}(t)$ derive from the real and imaginary parts
of~(\ref{phi1}), with the real part entering via
${\rm Im}(\phi_1^2)=2{\rm Re}(\phi_1)\,{\rm Im}(\phi_1)$.

We now investigate the limits as $d\to2$ and $d\to4$. The former case was
considered in the QCD$_2$ sum rules of~\cite{QCD2}, the latter was the
subject of the classic QED calculation of K\"{a}ll\'{e}n and
Sabry~\cite{KS}.

As $d\to2$, we obtain
\begin{equation}
r_1(t)\to\frac{2}{tv},\quad r_2(t)\to\frac{1}{2tv},\quad
r_3(t)\to\frac{-1}{8t^2v^3},\quad
r_4(t)\to\frac{3}{2tv^2}+\frac{3}{4t^2v^3}\ln x\,,
\label{r1234}
\end{equation}
with
\begin{equation}
x\equiv\frac{1-v}{1+v}\,,\quad v\equiv\sqrt{1-1/t}\,.
\label{xdef}
\end{equation}
The spectral density is obtained, from~(\ref{rho1},\ref{rho2}), as
\begin{equation}
\left.t^3\rho(t)\right|_{d=2}=\frac{1}{2v}\left(\frac{\alpha}{4\pi}\right)+
\Bigg\{\frac{1+v^2}{2v^3}+\frac{(1-v^2)^2}{2v^4}\ln x\Bigg\}
\left(\frac{\alpha}{4\pi}\right)^2+O(\alpha^3)\,.
\label{rh2}
\end{equation}
Making the transformations $z\to t+\ri0$, $y\to-x+\ri0$, $\ln y\to\ln
x+\ri\pi$, in~(\ref{phi}), we verify that~(\ref{rh2}) corresponds to the
discontinuity of the $d=2$ result of~(\ref{Pi12},\ref{Pi22}). Transforming
to the bare charge and mass, we verify the QCD$_2$ result of~\cite{QCD2}.
Since $\rho(t)$ has a non-integrable singularity at $t=1$, the dispersion
relation should be written for $(z-1)\Pi(z)$. The corresponding inclusion
of an additional factor of $(t-1)$ in the spectral integral will not
destroy its convergence at large $t$.

To obtain the $d=4$ spectral density of~\cite{KS}, we need to expand the
hypergeometric functions of~(\ref{r2}--\ref{r4}) to first order
in $\ep\equiv2-d/2$. The required expansions are
\begin{eqnarray}
&&\!\!\!\!\!\Fh21\Fft{\frac{d-1}{2},\frac12}{\frac{3d-5}{2}}=
\frac{15\sqrt{x}}{(1-x)^5}\Bigg\{
(1-x^2)(\Df14+x+\Df14x^2)\left[1-\Df{27}{10}\ep-4\ep\ln\frac{1+x}{2}\right]
\nonumber\\&&\quad\quad{}
+(x+x^2+x^3)\left[\ln x+\ep f(x)\right]
-\ep(\Df{31}{5}x+\Df{87}{10}x^2+\Df{51}{5}x^3+x^4)\ln x+O(\ep^2)\Bigg\},
\label{r2ex}\\
&&\!\!\!\!\!\Fh21\Fft{\frac{d-3}{2},\frac12}{\frac{3d-7}{2}}=
\frac{-3\sqrt{x}}{2(1-x)^3}\Bigg\{
(1-x^2)\left[1-3\ep-4\ep\ln\frac{1+x}{2}\right]
\nonumber\\&&\quad\quad{}
+(1+x^2)\left[\ln x+\ep f(x)\right]
-2\ep(3+x+5x^2)\ln x+O(\ep^2)\Bigg\},
\label{r3ex}\\
&&\!\!\!\!\!\Fh21\Fft{1,3-\frac{d}{2}}{\frac{7-d}{2}}
=\frac{-2x}{(1-x^2)}\Bigg\{(1+2\ep)\ln x-\ep g(x)+O(\ep^2)\Bigg\},
\label{r4ex}\\
&&f(x)=6\ln x\ln(1-x)-\Df12\ln^2x+6\Li2(x)-4\zeta(2)
+4\Li2(-x)+4\ln2\ln x,
\label{fx}\\
&&g(x)=2\ln x\ln(1-x)-\Df12\ln^2x+2\Li2(x)-2\zeta(2),
\label{gx}
\end{eqnarray}
from which we obtain
\begin{eqnarray}
\left.t\rho(t)\right|_{d=4}&=&
\frac{2v(3-v^2)}{3}\left(\frac{\alpha}{4\pi}\right)
+\Bigg\{2v(5-3v^2)+\frac{(1-v^2)(15+v^2)}{3}\ln x
\nonumber\\&&{}
+\frac{16v(3-v^2)}{3}
\left[\frac{1+v^2}{2v}+x\frac{\rd}{\rd x}\right]L_2(x)\Bigg\}
\left(\frac{\alpha}{4\pi}\right)^2+O(\alpha^3)\,,
\label{rh4}
\end{eqnarray}
in terms of the dilogarithmic discontinuity
\begin{eqnarray}
L_2(x)&\equiv&\left.m^2t\,
\frac{I_4\left(-q^2/4m^2\!=\!t+\ri0\right)
     -I_4\left(-q^2/4m^2\!=\!t-\ri0\right)}{2\pi\ri}\right|_{d=4}
\nonumber\\
&=&4\Li2(x)+2\Li2(-x)+\left[2\ln(1-x)+\ln(1+x)\right]\ln x\,,
\label{L2}
\end{eqnarray}
which was found by explicit integration in~\cite{BR} and has here been
obtained from the $O(\ep)$ terms of the hypergeometric
functions~(\ref{r2ex}--\ref{r4ex}). Differentiating~(\ref{L2}) in our
compact result~(\ref{rh4}), we easily reproduce the lengthier formul\ae\
of~\cite{KS,JS,BR}.

Here we have obtained the $d=4$ result of~(\ref{rh4},\ref{L2}) by
developing the $\ep$\/-expansion of the hypergeometric functions in our
algebraically computed $d$\/-dimensional result~(\ref{rho2}). This is the
generic method we propose for unsolved problems, such as the two-loop
gluon-condensate contributions to equal-mass current
correlators~\cite{GLUE}.

Note that our use of the regular basis $\{1,G_4,G_4^2,I_4,I_4^\prime\}$, to
obtain~(\ref{Pi24}) in the previous section, reduces the $d=4$ spectral
problem of this section to the calculation of the discontinuity of the
master integral~(\ref{i4q}), for $d=4$. Once this is obtained from
hypergeometric functions, as in~(\ref{L2}), all other spectral calculations
of this type are reduced to algebra. Once one has used Cauchy's theorem to
construct the function with this discontinuity, as was done in~\cite{DJB}
in the case of~(\ref{I4z}), all calculations of similar two-point functions
are reduced to $d$\/-dimensional algebra, amounting to no more than using
recurrence relations to find 5 rational coefficients that are regular as
$d\to4$.

\subsection{Accelerated convergence of the Taylor series}

For small $q^2/m^2=-4z$, we find the coefficients of the
Taylor series
\begin{equation}
\Pi_2(z)=\sum_{n=1}^\infty c(n)\left[-4z\right]^n\,,
\label{Pi2ex}
\end{equation}
from the hypergeometric series of~(\ref{phi1},\ref{phi2},\ref{phi3}) that
enter our $d$\/-dimensional result~(\ref{Pi2}). Running {\sc
Reduce}~3.4.1~\cite{RED} on a {\sc Vax}~4100, it took 20 seconds of CPUtime
to obtain the first 10 Taylor coefficients, the numerator of c(10) being a
polynomial in $d$ of degree 27. We list only
\begin{equation}
c(1)=\frac{4(3d^5-48d^4+227d^3-58d^2-1728d+1872)}
{9(d+2)d(d-3)(d-5)(d-7)},
\label{cc1}
\end{equation}
\begin{equation}
c(2)=\frac{11d^7-227d^6+1289d^5+1619d^4-31840d^3+30036d^2+263952d-311040}
{90(d+4)(d+2)d(d-3)(d-7)(d-9)},
\label{cc2}
\end{equation}
higher-order coefficients being too cumbersome to write here. The
$d\to4$ limit of the leading coefficient~(\ref{cc1}) gives the classic
result of~\cite{BDE}: $\frac14c(1)=-\frac{82}{81}+O(\ep)$.

Taking the $d\to4$ limit of the first 20 coefficients of~(\ref{Pi2ex}), we
find agreement with~\cite{SCG}, where a closed form is given for $d=4$. We
use these 4-dimensional coefficients to compute Pad\'{e} approximants to
the Taylor series. Computation of the two-loop spectral density reveals it
to be non-negative. Hence $\Pi_2$ is of Stieltjes type and convergence of
the Pad\'{e} approximants is guaranteed~\cite{EPS} throughout the region
$z\in[-\infty,1]$ and hence for arbitrarily large space-like $q^2$. Table~8
demonstrates the quality of the convergence. The notation is the same as in
Section 3.2: truncating the sum after $2n+1$ terms, with the first term
vanishing, we tabulate $E_4(n)$, the modulus of the relative error in the
4-dimensional $[n/n]$ Pad\'{e} approximant, for values of $q^2/m^2$ up to
12, well outside the disk $|q^2/m^2|<4$, to which the convergence of the
original Taylor series was restricted.

With just four non-vanishing terms in the Taylor series, the error $E_4(2)$
is less than $1\%$, for $q^2/m^2\leq10$. The inclusion of each new set of
four terms, reduces the error by at least two orders magnitude. Tables~6
and~8 clearly demonstrate the utility of applying Pad\'{e} approximants to
Taylor series, for obtaining accurate values of two-point functions, in the
space-like region. Since the calculation of the Taylor coefficients, from
vacuum diagrams, can be performed in the general mass case~\cite{INLO},
this is a most welcome result. We remark that the technology exists to
obtain {\em three\/}-loop QCD results by this method, since the calculation
of the relevant vacuum diagrams has been automated in the course of work on
three-loop terms in QED~\cite{3LP,BKT}. Methods of accelerated convergence
may also prove useful in the case of multi-loop diagrams with more than two
external particles, by improving expansions in powers of all the scalar
products of the external momenta.
\begin{table}[htb]
\caption{Accelerated convergence of the Taylor series of $\Pi_2$,
with $d=4$.}
\[\begin{array}{|c|llll|}\hline
q^2/m^2 &~~E_4(2)   &~~E_4(4)   &~~E_4(6)   &~~E_4(8)   \\ \hline
\Pm12.0 & \Pp112    & \Pp154    & \Pp196    & \Pp248    \\
\Pm11.0 & \Pp943    & \Pp114    & \Pp116    & \Pp128    \\
\Pm10.0 & \Pp783    & \Pp725    & \Pp637    & \Pp549    \\
\Pm 9.0 & \Pp633    & \Pp465    & \Pp327    & \Pp229    \\
\Pm 8.0 & \Pp493    & \Pp275    & \Pp147    & \Pp76{10} \\
\Pm 7.0 & \Pp363    & \Pp155    & \Pp578    & \Pp22{10} \\
\Pm 6.0 & \Pp253    & \Pp706    & \Pp188    & \Pp48{11} \\
\Pm 5.0 & \Pp163    & \Pp286    & \Pp459    & \Pp73{12} \\
\Pm 4.0 & \Pp894    & \Pp837    & \Pp73{10} & \Pp64{13} \\
\Pm 3.0 & \Pp394    & \Pp167    & \Pp60{11} & \Pp21{14} \\
\Pm 2.0 & \Pp114    & \Pp128    & \Pp13{12} & \Pp62{15} \\ \hline
\end{array}\]
\end{table}
\subsection{Accelerated convergence of the asymptotic expansion}

For large $q^2/4m^2=-z$, we invert the series
in~(\ref{phi1},\ref{phi2},\ref{phi3}) and obtain the coefficients of
\begin{equation}
\Pi_2(z)=\sum_{n=0}^\infty\frac{p(n)}{z^n}\,,
\label{Pi2asy}
\end{equation}
for arbitrary $d$. Each coefficient involves the structures $\beta$ and
$\gamma$, given in~(\ref{bet},\ref{gam}). In general, the terms
$\{1,\gamma,\beta,\beta^2\}$ occur, with rational coefficients that are
singular as $d\to4$. The $d\to4$ limit then gives a finite coefficient that
is quadratic~\cite{DJB} in $\ln(q^2/m^2)$. On the cut, with
$z\equiv-q^2/4m^2=t+\ri0$ and $t\in[1,\infty]$, one may extract the
asymptotic expansions of the real and imaginary parts of $\Pi_2(t+\ri0)$,
using $\ln(q^2/m^2)=\ln(4t)-\ri\pi$.
Here we list only the first three asymptotic coefficients, for arbitrary
$d$:
\begin{eqnarray}
\!\!p(0)&\!\!=&\!\!
-\frac{8(d^3-12d^2+41d-34)}{d(d-3)(d-4)(d-5)}
-\frac{32(d-2)(d^2-4d+8)}{(d-1)(d-4)^3(3d-8)(3d-10)}\,\gamma
\nonumber\\&&{}\!\!
+\frac{8(d-2)(d^2-7d+16)}{(d-1)(d-3)^2(d-4)^3}\,\beta^2,
\label{p0}\\
\!\!p(1)&\!\!=&\!\!
\frac{16(d-2)(d^3-8d^2+21d-26)}{(d-1)(d-3)(d-4)^2(d-6)(3d-10)}\,\gamma
+\frac{8(d-1)}{(d-3)^2(d-4)}\,\beta
\nonumber\\&&{}\!\!
-\frac{8(d-2)(d^2-6d+11)}{(d-1)(d-3)^2(d-4)^2}\,\beta^2,
\label{p1}\\
\!\!p(2)&\!\!=&\!\!
-\frac{32(d-1)}{d(d-2)(d-3)(d-4)^2}
-\frac{4(d-2)(d-5)(d^4-14d^3+65d^2-128d+128)}
{(d-1)(d-3)(d-4)^3(d-6)(d-8)}\,\gamma
\nonumber\\&&{}\!\!
-\frac{2(d^3-3d^2-26d+40)}{d(d-3)(d-4)^2}\,\beta
+\frac{4(d^3-10d^2+32d-36)}{(d-1)(d-4)^3}\,\beta^2.
\label{p2}
\end{eqnarray}

Table~9 shows accelerated convergence of the large-$q^2$ expansion of
$\Pi_2$, for $q^2/m^2\geq6$. The error $E_4(4)$ is at least three orders of
magnitude smaller than that which would have been obtained without
accelerating the convergence. As in Section~3.2, the application of the
$\ep$-algorithm, to both series, allows one to cover the entire space-like
region, to good accuracy, using only a few terms from each series.
Moreover, we find the $\ep$-algorithm to be similarly effective in
accelerating the convergence of the asymptotic expansions of the real and
imaginary parts on the cut.
\begin{table}[htb]
\caption{Accelerated convergence of the asymptotic expansion of $\Pi_2$,
with $d=4$.}
\[\begin{array}{|c|llll|}\hline
q^2/m^2 &~~E_4(1)   &~~E_4(2)   &~~E_4(3)   &~~E_4(4)   \\ \hline
\Pm40.0 & \Pp113    & \Pp977    & \Pp27{10} & \Pp23{13} \\
\Pm36.0 & \Pp143    & \Pp166    & \Pp51{10} & \Pp54{13} \\
\Pm32.0 & \Pp183    & \Pp276    & \Pp119    & \Pp14{12} \\
\Pm28.0 & \Pp253    & \Pp506    & \Pp249    & \Pp41{12} \\
\Pm24.0 & \Pp353    & \Pp996    & \Pp599    & \Pp14{11} \\
\Pm20.0 & \Pp533    & \Pp225    & \Pp178    & \Pp60{11} \\
\Pm16.0 & \Pp863    & \Pp585    & \Pp548    & \Pp33{10} \\
\Pm12.0 & \Pp162    & \Pp194    & \Pp177    & \Pp279    \\
\Pm10.0 & \Pp242    & \Pp364    & \Pp978    & \Pp639    \\
\Pm 8.0 & \Pp392    & \Pp654    & \Pp286    & \Pp498    \\
\Pm 6.0 & \Pp722    & \Pp114    & \Pp174    & \Pp207 \\ \hline
\end{array}\]
\end{table}
\section{Summary and conclusions}

There is a pressing need for techniques to calculate multi-loop radiative
corrections in processes with a wide variety of masses and momenta, in
order to probe the standard model. Here we have made a second step in
harnessing the economy of $d$\/-dimensional algebraic methods to that end.
The first step was made in~\cite{GBGS,SHELL2}, which dealt with on-shell
massive two-loop two-point functions. Here we have tackled all the
off-shell two-loop master integrals needed for QED and QCD, using
integration by parts to reduce all but one of them to hypergeometric
functions, and dealing with the troublesome fermion-propagator integral of
Fig.~1(c) equally effectively, by using differential equations, likewise
obtained from integration by parts.

As might have been expected from previous successes~\cite{GBGS,SHELL2,Q92}
of dimensional regularization of massive diagrams, our new
$d$\/-dimensional results are both easier to obtain and simpler in
structure than their 4-dimensional specializations. Just as the complexity
of polylogarithms~\cite{LEW} results from a singular case of hypergeometric
series, so 4-dimensional calculations are harder than $d$\/-dimensional
ones. The moral to be drawn from this is clear: take the limit $d\to4$ as
late as possible. The cost of carrying an extra variable in a
computer-algebra program is a small one to pay, for the avoidance of
analytic complexity, and can, in any case, be made even smaller, by
retaining only an appropriate number of powers of $(d-4)$.

We have applied algebraic methods to the two-loop photon self-energy. As
promised, a thorough knowledge of the analytic properties of the master
integral turns out to be sufficient to solve problems with arbitrary
spinorial or tensorial complexities, without need of further analysis. All
problems of the same mass structure (such as the problem~\cite{GLUE} of the
two-loop gluon-condensate contribution to vacuum polarization, which still
lacks a full solution) amount to no more than the determination of 5
rational functions of $d$ and $q^2/m^2$, multiplying structures that have
been thoroughly analyzed in Section~4.

We have shown examples in which the whole of the space-like region may be
accurately covered by applying Pad\'{e} approximants to the small-$q^2$
Taylor series, and their generalization, via the
$\ep$-algorithm~\cite{EPS}, to the large-$q^2$ asymptotic expansion.
Tables~8 and~9 exhibit such accelerated convergence, in the case of
two-loop vacuum polarization.

In the time-like region, our methods are similarly effective. In the case
of vacuum polarization, accelerated convergence of the Taylor series, for
$-q^2\in[0,4m^2]$, is achieved by applying the $\ep$-algorithm to
truncations of~(\ref{Pi2ex}). On the cut, with $-q^2=s+\ri0$ and
$s\in[4m^2,\infty]$, accelerated convergence of the asymptotic expansions
of the real and imaginary parts is achieved by applying the $\ep$-algorithm
to truncations of~(\ref{Pi2asy}). Thus we can approximate the two-loop
vacuum polarization function~(\ref{Pi24}) to good accuracy, throughout the
entire complex plane, using only a few of its algebraically computed Taylor
and asymptotic coefficients. Even near the branchpoint, with
$-q^2\approx4m^2$, we find the $\ep$-algorithm to be effective in
accelerating the convergence of the original series.

These efficient methods of algebraic computation and numerical
approximation are now available for {\em all\/} the self-energy diagrams of
QED and QCD, in {\em all\/} kinematical regions, thanks to our explicit
large and small $q^2$ expansions, in~(\ref{i4m},\ref{i2m},\ref{i3tm})
and~(\ref{i4q},\ref{i2q},\ref{i3tq}), and our analysis of the remaining
fermion-propagator integral of Section~2.5, with contributions that were
conveniently separated in Section~3.1. Given this separation, one may
accurately approximate the real and imaginary parts in the demanding
time-like region $-q^2=s+\ri0$, with $s\in[m^2,9m^2]$, by applying the
$\ep$-algorithm separately to the asymptotic expansion of the contribution
with the lower threshold and to the Taylor series of the contribution with
the higher threshold, paying attention to numerical cancellations in the
latter's coefficients, as in~(\ref{d20},\ref{das}). Elsewhere, one need not
make this delicate separation. Tables~6 and~7 of Section~3.2 show how good
is the coverage of the entire region $q^2\in[-m^2,\infty]$, thanks to the
enlarged domain of convergence of each series. The application of such
methods to the more general mass cases of~\cite{INLO,DST} should help to
extend the scope, speed and accuracy of standard-model calculations.

It now requires only careful algebraic programming to reduce the bosonic
and fermionic two-loop propagators of QED and QCD, in an arbitrary
covariant gauge, to the minimal set of master integrals of Fig~1, together
with a few of their derivatives. The application of methods of accelerated
convergence to the expansions of these propagators in powers of $q^2/m^2$
and $m^2/q^2$ will then yield economical and accurate parametrizations,
throughout the entire complex plane. Moreover, terms of order $(d-4)^{L-2}$
may also be obtained, for use in the dimensional regularization of
higher-order calculations, with $L$ loops.

It seems most likely that some degree of the analytical economy and
numerical accuracy achieved in this paper is also achievable in wider
areas, involving more loops, or more masses, or more external particles. We
plan to study three-loop diagrams.

\vspace{0.5cm}\noindent
{\em Acknowledgments\/}: D.J.B.\ thanks Vladimir Smirnov, for
discussion of the work in~\cite{DST}, and Ian McDonald, for assistance
in networking. O.V.T.\ is grateful to the Physics
Department of Bielefeld University, for warm hospitality, and to BMFT, for
financial support.

\newpage
\raggedright


\begin{thebibliography}{99}

\bibitem{LEP}
The LEP Collaborations, ALEPH, DELPHI, L3 and OPAL:
{\sl Phys.\ Lett.}\ {\bf B276} (1992) 247

\bibitem{STATUS}
L.\ Rolandi:
CERN--PPE--92--175, Precision tests of the electroweak interaction,\\
talk at ICHEP 92 conference, Dallas (August 1992);\\
G.\ Altarelli:
CERN--TH.6525/92, Precision electroweak data and constraints on new physics,
talk at XXVIIth Recontres de Moriond, Les Arcs (March 1992);\\
P.\ Renton:
{\sl Z.\ Phys.}\ {\bf C56} (1992) 355;\\
M.\ Cveti\v{c} and P.\ Langacker:
Testing the Standard Model (World Scientific, Singapore, 1991)

\bibitem{KK}
A.\ L.\ Kataev and V.\ T.\ Kim:
Annecy preprint ENSLAPP--A--407/92 (1992)

\bibitem{DJB}
D.\ J.\ Broadhurst:
{\sl Z.\ Phys.}\ {\bf C47} (1990) 115

\bibitem{BAK}
B.\ A.\ Kniehl:
{\sl Nucl.\ Phys.}\ {\bf B347} (1990) 86;\\
F.\ Halzen, B.\ A.\ Kniehl and M.\ L.\ Stong:
{\sl Z.\ Phys.}\ {\bf C58} (1993) 119

\bibitem{GBGS}
N.\ Gray, D.\ J.\ Broadhurst, W.\ Grafe and K.\ Schilcher:
{\sl Z.\ Phys.}\ {\bf C48} (1990) 673;\\
D.\ J.\ Broadhurst, N.\ Gray, and K.\ Schilcher:
{\sl Z.\ Phys.}\ {\bf C52} (1991) 111

\bibitem{SHELL2}
J.\ Fleischer and O.\ V.\ Tarasov:
{\sl Comp.\ Phys.\ Comm.}\ {\bf 71} (1992) 193

\bibitem{Q92}
D.\ J.\ Broadhurst:
OUT--4102--40 (1992), to appear in Proceedings of QUARKS--92

\bibitem{AI92}
D.\ J.\ Broadhurst:
in {\sl New computing techniques in physics research II\/},\\
ed.\ D.\ Perret--Gallix (World Scientific, Singapore, 1992) p.\ 579;\\
G.\ Weiglein, R.\ Mertig, R.\ Scharf and M.\ B\"{o}hm:
{\em ibid} p.\ 617;\\
J.\ Fujimoto, Y.\ Shimizu, K.\ Kato and Y.\ Oyanagi:
{\em ibid} p.\ 625

\bibitem{DK}
D.\ Kreimer:
{\sl Phys.\ Lett.}\ {\bf B273} (1991) 277;
Mainz preprint MZ--TH--92--50 (1992)

\bibitem{INLO}
A.\ I.\ Davydychev and J.\ B.\ Tausk:
{\sl Nucl.\ Phys.}\ {\bf B397} (1993) 123

\bibitem{DST}
A.\ I.\ Davydychev, V.\ A.\ Smirnov and J.\ B.\ Tausk:
INLO--PUB--5/93 (1993)

\bibitem{CT}
K.\ G.\ Chetyrkin and F.\ V.\ Tkachov:
{\sl Nucl.\ Phys.}\ {\bf B192} (1981) 159;\\
F.\ V.\ Tkachov:
{\sl Phys.\ Lett.}\ {\bf B100} (1981) 65

\bibitem{SVZ}
M.\ A.\ Shifman, A.\ I.\ Vainshtein and V.\ I.\ Zakharov:
{\sl Nucl.\ Phys.}\ {\bf B147} (1979) 385, 448, 519

\bibitem{GLUE}
P.\ A.\ Baikov et al:
{\sl Phys.\ Lett.}\ {\bf B263} (1991) 481;\\
K.\ G.\ Chetyrkin et al:
{\sl Phys.\ Lett.}\ {\bf B225} (1989) 411

\bibitem{KS}
G.\ K\"{a}ll\'{e}n and A.\ Sabry:
{\sl K.\ Dan.\ Vidensk.\ Selsk.\ Mat.-Fys.\ Medd.}\ {\bf 29} (1955) No.\ 17

\bibitem{JS}
J.\ Schwinger:
Particles, sources and fields
(Addison--Wesley, Reading, Mass., 1973) Vol.\ 2, p.\ 407

\bibitem{BR}
R.\ Barbieri and E.\ Remiddi:
{\sl Nuovo Cimento} {\bf 13} (1973) 99

\bibitem{PLB}
D.\ J.\ Broadhurst:
{\sl Phys.\ Lett.}\ {\bf B101} (1981) 423

\bibitem{SCG}
S.\ C.\ Generalis:
Open University thesis OUT--4102--13 (1984);\\
{\sl J.\ Phys.} {\bf G15} (1989) L225; {\bf G16} (1990) 367, 785, L115

\bibitem{QCD2}
P.\ Ditsas and G.\ Shaw:
{\sl Nucl.\ Phys.}\ {\bf B229} (1983) 29;\\
A.\ Bradley, C.\ S.\ Langensiepen and G.\ Shaw:
{\sl Phys.\ Lett.}\ {\bf B102} (1981) 180, 359

\bibitem{LEW} L.\ Lewin:
Polylogarithms and associated functions (North Holland, New York, 1981)

\bibitem{EPS}
D.\ Shanks:
{\sl J.\ Math.\ Phys.}\ {\bf 34} (1955) 1;\\
P.\ Wynn:
{\sl Math.\ Comp.}\ {\bf 15} (1961) 151;\\
G.\ A.\ Baker and P.\ Graves--Morris:
Pad\'{e} approximants,\\
in {\sl Encyclopedia of mathematics and its applications\/},\\
ed.\ G.-C.\ Rota (Addison--Wesley, Reading, Mass., 1981) Vol.\ 13, pp.\ 76--78

\bibitem{KOT}
A.\ V.\ Kotikov:
{\sl Phys.\ Lett.}\ {\bf B254} (1991) 185;
{\bf B259} (1991) 314;
{\bf B267} (1991) 123

\bibitem{JMP}
A.\ I.\ Davydychev:
{\sl J.\ Math.\ Phys.}\ {\bf 32} (1991) 1052;
{\bf 33} (1992) 358;\\
E.\ E.\ Boos and A.\ I.\ Davydychev:
{\sl Teor.\ Mat.\ Fiz.}\ {\bf 89} (1991) 56

\bibitem{3LP}
D.\ J.\ Broadhurst:
{\sl Z.\ Phys.}\ {\bf C54} (1992) 599

\bibitem{FJ}
C.\ Ford and D.\ R.\ T.\ Jones:
{\sl Phys.\ Lett.}\ {\bf B274} (1992) 409;
{\bf B285} (1992) 399(E);\\
C.\ Ford, I.\ Jack and D.\ R.\ T.\ Jones:
{\sl Nucl.\ Phys.}\ {\bf B387} (1992) 373;\\
E.\ Mendels:
{\sl Nuovo Cimento} {\bf A45} (1978) 87

\bibitem{RED}
A.\ C.\ Hearn:
REDUCE user's manual, version 3.4, Rand publication CP78 (1991);\\
J.\ A.\ M.\ Vermaseren:
Symbolic manipulation with FORM (Computer Algebra\\
Nederland, Amsterdam, 1991)

\bibitem{BDE}
M.\ Baranger, F.\ J.\ Dyson and E.\ E.\ Salpeter:
{\sl Phys.\ Rev.}\ {\bf 88} (1952) 680

\bibitem{BKT}
D.\ J.\ Broadhurst, A.\ L.\ Kataev and O.\ V.\ Tarasov:
{\sl Phys.\ Lett.}\ {\bf B298} (1993) 445

\end{thebibliography}
\end{document}